\documentclass[12pt]{article}%
\usepackage{heck}
\usepackage{cite}
\usepackage{graphicx}
\usepackage{makeidx}
\usepackage{multicol}
\usepackage{geometry}
\usepackage{amsfonts}
\usepackage{mathrsfs}
\usepackage{amssymb}
\usepackage{amsmath}%

\providecommand{\U}[1]{\protect\rule{.1in}{.1in}}
\numberwithin{equation}{section}

\newcommand{\bea}{\begin{eqnarray}}
\newcommand{\eea}{\end{eqnarray}}
\newcommand{\be}{\begin{equation}}
\newcommand{\ee}{\end{equation}}

\newcommand{\bem}{\begin{pmatrix}}
\newcommand{\eem}{\end{pmatrix}}

\def\a{\alpha}

\def\d{\delta}

\def\e{\epsilon}
\def\f{\phi}               

\def\inf{\infty}

\def\k{\kappa}             

\def\m{\mu}

\def\o{\omega}  
\def\ost{\circledast}
\def\p{\pi}
\def\pa{\partial}
\def\r{\rho}                                     
\def\t{\tau}
\def\th{\theta}
\def\til{\tilde}

\def\z{\zeta}

\def\L{\Lambda}
\def\O{\Omega}

\def\Tr{{\rm Tr}}
\def\U{\Upsilon}

\def \Z {{\mathbb Z}}
\def \C {{\mathbb C}}
\def \R {{\mathbb R}}
\def\fnull{\varphi^{(0)}}
\def\fone{\varphi^{(1)}}
\def\finull{\phi^{(0)}}
\def\fione{\phi^{(1)}}
\def\anull{\bar{A}^{(0)}}
\def \aone{\bar{A}^{(1)}}

\begin{document}

\date{October, 2009}

\preprint{arXiv:0910.0477}

\institution{SISSA}{\centerline{${}^{1}$Scuola Internazionale Superiore di Studi Avanzati via Beirut 2-4 I-34100 Trieste, ITALY}}

\institution{HarvardU}{\centerline{${}^{2}$Jefferson Physical Laboratory, Harvard University, Cambridge, MA 02138, USA}}%

\institution{IAS}{\centerline{${}^{3}$School of Natural Sciences, Institute for Advanced Study, Princeton, NJ 08540, USA}}

\title{Yukawa Couplings in F-theory and Non-Commutative Geometry}
%

\authors{Sergio Cecotti\worksat{\SISSA}\footnote{e-mail: {\tt cecotti@sissa.it}}, Miranda C.N. Cheng\worksat{\HarvardU}\footnote{e-mail: {\tt mcheng@physics.harvard.edu}}, \\[2mm]Jonathan J. Heckman\worksat{\IAS}\footnote{e-mail: {\tt jheckman@sns.ias.edu}} and Cumrun Vafa\worksat{\HarvardU}\footnote{e-mail: {\tt vafa@physics.harvard.edu}}}%

\abstract{We consider Yukawa couplings generated by a configuration of intersecting seven-branes in F-theory. In configurations with a single interaction point and no fluxes turned on, the Yukawa matrices have rank one.
This is no longer true when the three-form $H$-flux is turned on, which is generically the
case for F-theory compactifications on Calabi-Yau fourfolds.
In the presence of $H$-fluxes, the Yukawa coupling is computed using a
non-commutative deformation of holomorphic Chern-Simons theory (and its reduction to seven-branes) and subsequently the rank of the Yukawa matrix changes.
Such fluxes give rise to a hierarchical structure in the Yukawa matrix in
F-theory GUTs of the type which has recently been proposed as
a resolution of the flavor hierarchy problem.}%

\maketitle

\tableofcontents

\pagebreak

\section{Introduction}

It has been known for a long time that the existence of Yukawa couplings is crucial in understanding the masses and mixing angles
for quarks and leptons. Indeed, Yukawa matrices in flavor space determine the structure of the corresponding mass
matrices. These parameters exhibit striking hierarchies which are not explained by
the Standard Model. Turning this around, the existence of hierarchical structure in the
Yukawa couplings can serve as a hint for physics beyond the Standard Model. It is thus
natural for string theory to use this hint as a way to connect to particle phenomenology.

This idea is very natural in the context of ${\cal N}=1$ supersymmetric theories, since
Yukawa couplings are holomorphic and as such are often exactly computable.\footnote{Of course the
physical Yukawa couplings also depend on the normalization of fields which are not holomorphic quantities,
and are therefore hard to compute exactly.} There are various contexts in which Yukawa couplings
appear in string theory, and in all cases these computations are quasi-topological
in nature.  Assuming that supersymmetry is a part of Nature at some scale, this suggests a
link between hierarchical structures in the internal geometry, and those of the Standard Model.

In particular, it is well-known that up-type quarks have a very hierarchical mass structure. The top quark is very heavy, with a
mass of $\sim 170$ GeV, which is close to that of the weak scale, while the up and charm are far lighter. It is
therefore natural to search for a topological scenario in string theory where we start with one massive
quark and two massless ones.  In other words, it is natural to approximate the Yukawa matrix for the (u,c,t) quarks as a
rank one matrix which is then deformed by higher order hierarchical corrections.

Recent work has shown that a class of Grand Unified Theories in F-theory known as F-theory GUTs
provide a potentially promising starting point for making detailed contact between string theory
and the Standard Model. See for example
\cite{DonagiWijnholt,beasleyheckmanvafa1,Aparicio:2008wh,WatariTATARHETF,Buchbinder:2008at,
BHVII,HMSSNV,MarsanoGMSB,DonagiWijnholtBreak,MarsanoToolbox,HVGMSB,
HVLHC,Font:2008id,Heckman:2008qa,Blumenhagen:2008zz,Blumenhagen:2008aw,FGUTSCosmo,Bourjaily:2009vf,tataretal,
Andreas:2009uf,Chen:2009me,HKSV,DonagiWijnholtIII,BHSV,Randall:2009dw,HVCP,
Bourjaily:2009ci,Tatar:2009jk,Jiang:2009za,Blumenhagen:2009up,
Heckman:2009mn,Font:2009gq,Marsano:2009gv,Blumenhagen:2009yv}
for recent work on F-theory GUTs. In these theories, matter and Yukawa interactions respectively localize on Riemann surfaces and at points of an intersecting seven-brane configuration. Yukawas are controlled by the overlap of matter field wave-functions and are concentrated at the point of maximal overlap. As observed in \cite{beasleyheckmanvafa1} (see also \cite{tataretal}), minimal geometries with a single Yukawa point generate rank one Yukawa matrices, and adding more points simply produces higher rank structures.

Building on this observation, it was proposed in \cite{Heckman:2008qa} that starting from a minimal geometry with a single Yukawa point,
subleading corrections to this geometric structure could generate hierarchical flavor structure for the remaining generations. Subleading corrections
can in principle be generated by the spread of the matter field wave-functions in the internal geometry.
Hierarchical structures are then expected from small flux induced distortions to the matter field
wave-functions. Furthermore, the $U(1)$ symmetries induced from local rotations of the
complex internal space at the intersection point constrain the structure of the hierarchical
corrections to the Yukawas, implementing a mechanism rather similar to
the Froggatt-Nielsen mechanism \cite{Froggatt:1978nt}.

A crucial element of this proposal is that the leading order deformation to the
Yukawas originates from a three tensor index structure, such as the gradient
of the gauge field strength $\nabla_{\bar{i}} F_{j\bar k}$ or $\nabla_i F_{\bar{j} \bar{k}}$. It was
argued in \cite{Heckman:2008qa} that the constant two-form flux $F$ on the
seven-brane alone will not induce any correction to the Yukawa couplings.
Using crude estimates for the expected form of flux-induced deformations, it
was found that hierarchies in the quark masses and mixing angles could naturally
be generated. One of the primary aims of this paper is to evaluate this proposal.

Recall that there are essentially two types of fluxes in F-theory backgrounds. These correspond
to gauge field fluxes $F$ threading the worldvolume of a seven-brane, and to bulk
three-form $H$-fluxes which can be decomposed as $H = H_{R} + \tau H_{NS}$ in perturbative IIB language.
In a dual M-theory description, these two contributions both descend from a four-form $G$-flux.\footnote{It
is also sometimes possible to consider F-theory vacua with background five-form fluxes activated. Such fluxes lie
somewhat outside the realm of the local model defined by a configuration of seven-branes, and so we will not
consider this case in this paper.} When {only} $F$ is activated, we find that no change can occur in
the rank of the Yukawa matrix away from the minimal form expected from geometry. We shall sometimes refer
to this result as the ``rank theorem'' for Yukawa couplings in commutative geometry. This result also implies that more general F-theory backgrounds must be considered to deform the structure of the Yukawa couplings. When
$H$-flux also participates, however, we find that the rank generically {does} increase,
and that the form of these subleading corrections generate hierarchical structures which
are almost identical to those proposed in \cite{Heckman:2008qa}. Indeed, in simplified situations,
the presence of the three-form flux $H$ can be viewed as a mild shift in the gradient of the background gauge field strength $\nabla F = -\nabla B$.

At a pragmatic level, it is possible to work in ``physical gauge'', by specifying a choice of background K\"{a}hler metric, computing the precise profile of the matter field wave-functions in this background and then evaluating the explicit
overlap integrals in a local neighborhood containing the Yukawa enhancement point. This was the point of view advocated for
example in \cite{Heckman:2008qa}. On the other hand, in a given overlap integral, there will typically be non-trivial cancellations from distinct terms. The primary advantage of working in terms of a manifestly holomorphic formulation is that such cancellations are immediate from the start of the computation. Indeed, when only the gauge flux $F$ is present, we find highly non-trivial cancellations in the form of the overlap integral which are quite subtle in physical gauge. In the present paper we have performed the relevant overlap integrals in both holomorphic and physical wave-function bases as a check on the expressions found.

The holomorphic formulation of Yukawa couplings is based on the observation made in \cite{beasleyheckmanvafa1} that
the Yukawa couplings in F-theory are computed using the reduction of holomorphic Chern-Simons theory to the seven-brane
world-volume. In other words, the matter fields correspond to first order deformation moduli
of the holomorphic Chern-Simons theory, and the Yukawa couplings measure the obstruction
to extending these moduli to higher order. We show how the computation of local Yukawa couplings
can be simplified and recast in the form of a residue formula. This residue formula for the Yukawas
turns out to be  powerful in establishing the rank theorem when $H=0$.

From the perspective of the quasi-topological theory which computes the coupling, the effects of $H\neq 0$ can be captured in terms of a non-commutative deformation of holomorphic Chern-Simons theory and its reduction to a configuration of seven-branes.\footnote{For other applications of non-commutative Chern-Simons theory, see for example \cite{Susskind:2001fb}.} In the non-commutative geometry, the holomorphic coordinates $X^{i}$ and $X^j$ satisfy
the relation:
$$[X^i,X^j]= \theta^{ij}(X)$$
where $\theta^{ij}$ is a holomorphic bi-vector. It follows from the Jacobi identity of this commutator that $\th^{ij}$ is a Poisson bi-vector:
$$\theta^{i\ell}\,\partial_\ell\,\theta^{jk}+({\rm antisymmetrization\ in\ } i,j,k)=0\;.$$
As we will argue, the non-commutativity parameter $\theta^{ij}$ is related to the $H$-flux by:
$$\nabla_k \theta^{ij}=H^{\,\;\;ij}_k$$
to leading order in the weak field expansion.

This kind of non-commutativity is encountered in the context of topological strings
\cite{kapustin,Kapustin:2004gv,deformation}.  Here we show that the natural setting in
the superstring context involves the $H$-flux.  It is well-known that
this kind of non-commutativity does not typically arise in perturbative string compactifications because the $H$-flux
satisfies a Gauss' law type constraint.\footnote{Compactifications on hyperk\"{a}hler manifolds and non-compact
Calabi-Yau threefolds provide exceptions to this general point, but also serve to illustrate just how strong this constraint is.}
On the other hand, it is also well known that F-theory vacua typically do have $H$-flux in compact
geometries \cite{Gukov:1999ya}. Indeed, F-theory vacua generically evade such no-go theorems because
the $SL(2,\mathbb{Z})$ duality group acts non-trivially on $H$. At a practical level, what this means is that far
from being an exotic element of a compactification, background $H$-fluxes will generically deform the structure of
F-theory Yukawas in potentially phenomenologically interesting ways.

The resulting hierarchical structures we find are quite similar to those initially proposed in \cite{Heckman:2008qa}. Indeed, another benefit of a manifestly holomorphic formulation is that it is possible to give more reliable estimates on the size of the ``order one'' coefficients expected in each entry of the Yukawa matrix. We find that in all examples we consider, the magnitudes of the coefficients multiplying each hierarchical suppression factor are indeed order one numbers, in accord with crude numerological expectations.

The organization of the rest of this paper is as follows.
In section 2 we review aspects of F-theory in connection with Yukawa
couplings.  In section 3 we show how one can compute the Yukawa
couplings for seven-branes locally in terms of a residue formula.
We also derive the rank theorem which is valid in the absence of  $H$-fluxes in this section. In section 4 we show why $H$-fluxes are generic in F-theory compactifications on Calabi-Yau fourfolds and why the F-terms are computed by non-commutative holomorphic Chern-Simons theory.  In section 5 we
consider non-commutative holomorphic Chern-Simons theory and its reduction to seven-branes. In particular, we
compute the Yukawa couplings in this setup, finding hierarchical corrections to the leading order result. In section 6 we apply our
results to the study of flavor hierarchy and recover structures very similar to those found in \cite{Heckman:2008qa}.  Section 7 contains
our conclusions and future directions of investigation. Appendix A is devoted to further elaboration on the residue formulation of Yukawas in commutative geometries, and Appendices B and C contain some technical aspects of the
non-commutative deformation and its application to deformed F-terms.
\section{F-theory Yukawas and Seven-Branes}
\label{F-theory Yukawas and Seven-Branes}

The primary aim of this paper is to study the superpotential of a configuration of
seven-branes in the presence of non-trivial background fluxes in F-theory. In this
section we recall the main ingredients entering into the computation of Yukawa couplings in F-theory vacua.

F-theory can be defined as a strongly coupled formulation of type IIB string theory in which
the axion-dilaton is allowed to vary over points of the ten-dimensional spacetime. This information
can be conveniently packaged in terms of a twelve-dimensional geometry. F-theory compactified
on an elliptically fibered Calabi-Yau fourfold
with a section preserves $\mathcal{N} = 1$ supersymmetry in the four uncompactified dimensions.
The discriminant locus of the elliptic fibration defines the location of seven-branes
wrapping complex surfaces in the threefold base $X$ of the elliptic fibration.
Locally, the seven-brane wrapping a complex surface $S$ can be described as a
singularity of ADE type fibered over $S$. Six-dimensional chiral matter fields correspond to the enhancements
of the singularity type to higher rank along complex curves, and Yukawa interactions correspond
to further enhancements of the singularity type at points in the geometry.

In addition to the geometric elements of an F-theory compactification, there can
also be additional contributions from background fluxes. Such fluxes can either descend from two-form fluxes, such as gauge field fluxes on the various seven-branes, or from three-form NSNS and RR fluxes (in type IIB language).
Upon further compactification on a circle down to three dimensions, we have the dual M-theory description on the same Calabi-Yau fourfold, where the K\"ahler class of the elliptic fiber is the inverse of the radius. It is well-known that in this context, one generically has a (2,2) $G_4$-flux turned on \cite{Becker:1996gj}. Lifting it back up to F-theory, this corresponds to turning on  $H$-flux \cite{Sethi:1996es,Gukov:1999ya}:
\begin{equation}\label{G4andH}
G_{4} = H \wedge dz + \overline{H} \wedge d \bar{z}\;,
\end{equation}
where $z$ is the holomorphic coordinate along the elliptic fiber of the Calabi-Yau fourfold, and $H=H_{R} + \tau H_{NS}$ is a (1,2)-form. As
reviewed for example in \cite{Denef:2008wq}, abelian gauge field fluxes threading the seven-brane correspond to an appropriate localization of the $G_4$-fluxes on the components of the discriminant locus, while the remaining degrees of freedom of the $G_4$-fluxes correspond to the bulk three-form fluxes $H$ of interest to us in the present paper.

The background fluxes of an F-theory compactification will in general distort the profile of matter field wave-functions of the compactification. At an abstract level, these chiral matter fields correspond to representatives of elements in an appropriately defined cohomology theory. Indeed, the zero mode wave-functions $\Psi^{(i)}$ localized on a matter curve $\Sigma$ are defined in terms of the Dirac equation:
\begin{equation}
\bar{D} \Psi^{(i)} = 0
\end{equation}
where $\bar{D}$ is an appropriate Dolbeault operator defined in terms of the background fluxes of the compactification, and we have
labelled the distinct zero modes by the index $i=1,...,n$ for $n$ zero modes localized on the curve $\Sigma$.

Yukawa couplings between three matter fields in the four-dimensional effective
theory are given by triple overlaps of wave-functions in the internal
directions of the geometry:%
\begin{equation}
\lambda^{ijk}=%
{\displaystyle\int}
\Psi_{\Sigma_{1}}^{(i)}\Psi_{\Sigma_{2}}^{(j)}\Psi_{\Sigma_{3}}^{(k)}.
\label{overlapYuk}%
\end{equation}
For phenomenological applications, the case of primary interest is where one of the matter field wave-functions corresponds to a Higgs field of the MSSM. For this reason, we shall be primarily interested in cases where the Yukawa couplings reduce to a $g \times g$ matrix for a $g$-generation model.

Although this might suggest that it is necessary to specify the explicit form of the matter wave-functions, much of this data is unnecessary in a computation of F-terms due to holomorphy considerations. Specifically, note that for any appropriately defined theory of Yukawa couplings, the corresponding overlap integral should be independent of the particular representative chosen for the matter wave-function. Indeed, this is a necessary property of the superpotential in order for it to have a formulation which is independent of the D-term data, such as the K\"ahler form of the complex surface.\footnote{Here we are neglecting possible instanton type contributions to the superpotential which contain terms of the form $\exp(-\text{Vol}(S))$.}

A more pragmatic benefit of having a holomorphic formulation of Yukawa couplings is that it also provides a useful check that there are no non-trivial cancellations between different contributions to the wave-function overlap integrals. Indeed, although such cancellations can appear to be rather mysterious and non-trivial in ``physical gauge'', holomorphy considerations typically make such effects more transparent.

In the context of perturbative type IIB string theory, the evaluation of Yukawas reduces to a disc diagram computation in the topological B-model. Phrased in this way, our primary task is to determine the deformation theory of the Yukawa couplings in the topological B-model. Once these deformations have been determined, we next establish a link to the physical theory. In other words,
it is then enough to determine how various background field configurations in the physical theory such as fluxes translate into parameters in the topological string.

Recall that in the topological B-model on a Calabi-Yau threefold $X$, the closed string moduli space is:
\begin{equation}
\underset{p,q}{\oplus} H^{p}(\Lambda^{q}T^{1,0})
\end{equation}
where $T^{1,0}$ denotes the holomorphic tangent bundle. This moduli space is known as the ``extended moduli space'' in the sense of Witten \cite{Witten:1991zz} (see also \cite{deformation}). The more familiar class of deformations satisfying $ p+q = 2$ correspond to marginal deformations of the worldsheet theory. For example, $p=q=1$ correspond to the usual complex deformations typically studied in the context of the B-model. Here, the case of $p = 2$ and $q = 0$ corresponds to ``gerby'' deformations, and $p = 0$ and $q = 2$ corresponds to deformations which induce the Kontsevich $\star$-product between open string states \cite{deformation}.

On the other hand, one of the most attractive features of F-theory compactifications is that some Yukawas which cannot be realized perturbatively are naturally generated from E-type singularity structures in the fibration of the elliptic Calabi-Yau fourfold. Even so, it is still possible to retain computational control over some aspects of the seven-brane theory by appealing to the formulation of the partially twisted eight-dimensional gauge theory \cite{beasleyheckmanvafa1}. Since Yukawa couplings correspond to points of the geometry where the singularity type enhances to $G_{p}$, in a sufficiently small neighborhood of this point, the physical system defined by this geometry is given by a gauge theory with gauge group $G_{p}$ which is Higgsed down to $G_{\Sigma}$ along matter curves, and $G_{S}$ in the bulk of the complex surface wrapped by the seven-brane. This provides a topological formulation of Yukawa couplings in a patch containing the enhancement point, and this is the approach we shall follow in this paper. Furthermore, extending the relationship between deformations of the topological theory by type IIB field configurations to F-theory, we will be able to study the Yukawa couplings  in the partially twisted eight-dimensional gauge theory deformed by the presence of F-theory fluxes.

\section{Yukawas in Commutative Geometry}
\label{Yukawas in Commutative Geometry}

Our main aim in this section will be to derive a quasi-topological
formulation of Yukawa couplings among the localized zero modes of
the theory. The formulation of Yukawa couplings we present will be
based on the local profile of the background fields in a sufficiently
small neighborhood of the Yukawa interaction point. For this reason,
it is often sufficient to study the gauge theory as if it were defined
in a patch of $\mathbb{C}^2$. Within this patch, we shall model the
system as an eight-dimensional gauge theory with gauge group $G_p$ which away from the Yukawa point has been Higgsed down to $G_{\Sigma}$ along matter curves, and $G_{S}$ in the bulk.

A quasi-topological formulation of Yukawas helps to illustrate two important
features. First, the reduction of the computation to
a residue integral demonstrates that small changes in the shape
of the patch cannot alter the value of the Yukawa coupling, and hence the Yukawa
coupling is a well-defined observable of the quasi-topological theory. Second,
in this manifestly holomorphic formulation, it is straightforward to demonstrate
that background gauge field fluxes do not alter the structure of the Yukawa
couplings.

This result may appear surprising at first sight since background gauge fields do distort the profile of the matter field wave-functions.
As a check on these results, we have also computed in some explicit examples presented in section \ref{Two Examples}
the same overlap integrals by using wave-functions which satisfy both the F-term and D-term equations of motion. In
this ``physical gauge'', the holomorphy of the superpotential is more obscure, but the actual profile of the matter field wave-functions
is easier to track. The absence of corrections to the Yukawa couplings can be traced to
subtle cancellations between different contributions to the wave-function overlap integrals. It follows
from these computations that additional ingredients must be included in the seven-brane theory if we are to realize
flavor hierarchies in F-theory GUTs. In subsequent sections we shall study the hierarchies generated by one such deformation.

The rest of this section is organized as follows. After a quick review of the seven-brane gauge theory, we discuss the notion of matter curves and zero modes localized on them. We then show how, for a given background, the gauge-inequivalent localized zero modes are captured by a cohomology theory. From there we derive a residue formula for the Yukawa triple coupling of the localized zero modes and arrive at a theorem on the rank of the Yukawa matrix. As checks of the formula, we also present two explicit examples in which we calculate the zero mode wave-functions for a given choice of K\"ahler form and background gauge flux. We find that these computations agree with the results of the residue formula.

\subsection{The Partially Twisted Seven-Brane Theory}

In this subsection we will briefly review the partially twisted supersymmetric Yang-Mills theory describing seven-branes filling the Minkowski spacetime $\R^{3,1}$ and wrapping a K\"ahler surface $S$ \cite{beasleyheckmanvafa1} (see also \cite{DonagiWijnholt}).

To motivate this theory, recall that the superpotential of the field theory describing a stack of D9-branes with gauge group $G$ wrapping a Calabi-Yau threefold $X$ is determined by the holomorphic Chern-Simons Lagrangian \cite{Witten:1992fb}
\be
 \int_X \O^{3,0} \wedge\Tr \big( \bar A \wedge \bar\pa \bar A +\frac{2}{3} \bar A \wedge \bar A \wedge \bar A  \big) \;
\ee
with $\bar A$ the connection for $G$. Here and in what follows we have suppressed the non-compact $\R^{3,1}$ directions which do not play any role in our discussions. Given a background gauge field configuration $\langle {\bar A} \rangle$, zero mode fluctuations $\delta {\bar A}$ about this background correspond to massless excitations of the four-dimensional effective theory.

The superpotential of the seven-brane gauge theory can be defined by dimensional reduction of the holomorphic Chern-Simons theory. Schematically, denoting by ${\bar A}_{\bar\perp}$ the component of the gauge field transverse to the seven-brane world-volume and replacing $\O_{ij\perp} {\bar A}_{\bar \perp}$ by $\varphi_{ij}$, the superpotential for a seven-brane with gauge group $G$ on $\R^{3,1}\times S$ is given by the supersymmetrization of:
\be\label{superpotential}
W = \int_S \Tr \big(\varphi \wedge F^{(0,2)}\big) \;,
\ee
where $F^{(0,2)}$ denotes the $(0,2)$-component of the gauge field strength:
$$
F^{(0,2)} = {\bar \partial} {\bar A} + {\bar A} \wedge {\bar A}.
$$
When the context is clear, we shall drop the label indicating the Hodge type of the form. In addition, we shall not distinguish between chiral superfields and their bosonic components given that there is little room for confusion. Notice from the above prescription that we can naturally identify $\varphi$ as a $(2,0)$-form on the complex surface $S$. This
is consistent with the result obtained from the unique twisting of ${\cal N}=1$
supersymmetric Yang-Mills theory on $\R^{3,1}\times S$ when $S$ is a K\"ahler surface
\cite{beasleyheckmanvafa1}. Let us note here that although we have motivated this
superpotential using a formulation based on dimensional reduction and compactification
on a Calabi-Yau threefold, all that is really required to derive the superpotential
of the seven-brane theory is the existence of $\mathcal{N} = 1$ supersymmetry in
$\mathbb{R}^{3,1}$. Indeed, this is the original approach used in \cite{beasleyheckmanvafa1}.
This approach is also useful in cases which cannot be realized in perturbative D-brane constructions.

From the above superpotential it is straightforward to derive the F-term equations of motion
\bea\label{ftermeq1}
F^{(0,2)} &=& \bar \pa \bar A + \bar A\wedge \bar A=0 \\
\label{ftermeq2}
\bar\pa_A \varphi &=& 0\;.
\eea
The first equation ensures $\bar \pa_A^2 = 0$ and hence the existence of a complex
structure on the gauge bundle, while the second equation states that the
adjoint-valued two-form field $\varphi$ is holomorphic with respect to this complex structure.

The superpotential of the seven-brane theory is invariant under the gauge transformations:
\be\label{gauge_sym}
\bar A \to g^{-1} \bar A \,g + g^{-1} \bar \pa  g\quad,\quad \varphi \to g^{-1}  \varphi\, g\;.
\ee
Note that in the present theory the action is invariant not only under real gauge transformations, but also complex (or holomorphic) gauge transformations.
As we will see later, this gauge invariance will allow us to develop a cohomology theory for the zero mode solutions and will also greatly simplify our computation of the Yukawa couplings between chiral matter fields.

Up to this point, our discussion has focussed solely on the superpotential of the seven-brane theory. The D-term equation of motion for the partially twisted seven-brane is given by \cite{beasleyheckmanvafa1}:
\be\label{d_term_1}
\o\wedge F + \frac{i}{2} [\varphi,\bar \varphi]=0 \;,
\ee
where $\o$ is the K\"ahler form on $S$. This D-term constraint constitutes non-holomorphic data,
and so is not expected to play a crucial role in a holomorphic formulation of the Yukawa
couplings. As we will explain later, in our case this can be explicitly
seen from the gauge invariance of the superpotential.

\subsection{Matter Curves and Localized Zero Modes}
\label{The Matter Curves and the Localized Zero Modes}

In this subsection we will first review part of \cite{beasleyheckmanvafa1},
showing how the intersection of other seven-branes with the seven-brane on $S$ on which our gauge theory is defined can be modelled by a specific background field configuration, and how chiral matter fields localized on the intersection locus are related to fluctuations around this background. After this we will see how the chiral modes can be described by a cohomology theory. In subsequent sections this will make manifest the quasi-topological nature of the Yukawa coupling.

\subsubsection{Enhancement Loci and Matter Curves}
\label{The Matter Curves}

Following \cite{Katz:1996xe,beasleyheckmanvafa1}, we now discuss how matter curves are specified by a background field configuration of the seven-brane theory. This amounts to a choice of supersymmetric background field configuration $\anull$, $\fnull$. Since the main object of interest in the present paper is the triple coupling of chiral zero modes localized on three ``matter curves'' meeting at one point, we can work in a local neighborhood $U$ of this point.

In a local patch, the BPS equation $\bar \pa_{A^{(0)}}^2=0$ ensures we can further simplify our discussion by working in the holomorphic gauge
\be \label{holomorphic_gauge}\anull=0\;.\ee As guaranteed by the gauge invariance of the theory, no generality is lost by making such a choice.

To discuss the background for the two-form field $\fnull$, recall that it is a section of the canonical bundle of the surface $S$. Adopting local holomorphic coordinates $x$ and $y$ in the patch $U$, we can simply write $dx\wedge dy$ as a basis for the $(2,0)$-form. Writing
$$
\fnull = \f^{(0)} dx\wedge dy\;,
$$
let us now focus on the simplest kind of background in which $\varphi^{(0)}$ takes values in the Cartan subalgebra $\mathfrak{h}$ of the Lie algebra $\mathfrak{g}$ of the gauge group $G_p$:
\be\label{cartan_bkgnd}
\finull \in \mathfrak{h}.\ee
In this case, the D-term equation (\ref{d_term_1}) reduces to the usual instanton equation $\o\wedge F=0$. The non-zero value for $\varphi^{(0)}$ specifies a breaking pattern for $G_{p}$.

A justification for this choice of $\fnull$ comes from the physical consideration that we are interested in modelling a situation in which other seven-branes intersect the surface $S$ along some curves $\Sigma$ and the gauge symmetry on $S$ is enhanced along the loci of intersection. For $\finull$ valued in the Cartan subalgebra $\mathfrak{h}$, the gauge group $G_p$ is broken to its maximal torus  at a generic point on $S$, as is the case for intersecting branes. At a generic point, no gauge symmetry will be left when $\finull$ is not in the Cartan.\footnote{It would be interesting to extend the analysis presented here to more general choices for $\phi^{(0)}$, such as cases where $\phi^{(0)}$ is nilpotent. See for example, \cite{Donagi:2003hh} for a study of the spectrum of B-branes in a similar setup.}

The general configuration of matter curves is conveniently packaged in terms of the root space decomposition of the Lie algebra $\mathfrak{g}$ of $G_{p}$ so that:
\begin{equation}
\mathfrak{g} = \mathfrak{h} \oplus_\alpha \mathfrak{g}_\a
\end{equation}
where $\alpha$ denotes the roots of the algebra. For $\finull \in \mathfrak{h}$ we can write
$$
[\finull,X_\a] = \a(\finull) X_\a\;,
$$
where $X_\a$ is the generator in the Lie algebra corresponding to the root $\a$. From this expression, it is now immediate that the curve
$\Sigma_\alpha$ defined by $\a(\finull)=0$ for some positive root $\alpha$ corresponds to a locus of gauge enhancement
and will hence be identified as the curve of intersection between seven-branes.

\vspace{18pt}
{\it{An Example}}
\vspace{12pt}

To illustrate the main ideas discussed above, we now consider a system with gauge group $U(1)$ at a generic point on the surface $S$, and a matter curve where the singularity type enhances to $SU(2)$.
In this case, we can choose the gauge and the coordinates such that our background configuration reads
$$
\anull=0\quad,\quad \fnull = \bem x &0\\ 0&-x\eem dx \wedge dy\;,
$$
where $x,y$ denote local coordinates of the patch. The locus of gauge
enhancement in these coordinates is the complex line $x=0$. Note that
in more general situations where $\fnull$ embeds as a $2 \times 2$ sub-block
of a larger (traceless) matrix, we can also consider configurations of the form:
$$
\anull=0\quad,\quad \fnull = \bem x + \Phi &0\\ 0&-x + \Phi\eem dx \wedge dy\;
$$
for $\Phi = \Phi(x,y)$. In this case, the matter curve defined as the gauge enhancement locus is still given by the complex line $x = 0$.

\subsubsection{Zero Modes Localized on Curves}
\label{Zero Modes Localized on Curves}

In order to study the chiral matter in a specific background, we first determine the available deformations of a given supersymmetric background field configuration. The massless fluctuations about this background will then define the zero modes of the system.

Separating the fields $\varphi$ and $\bar A$ into a background value and fluctuations about this background, we write:
\be\label{separate_bkgnd}
\varphi = \fnull+ \fone\quad,\quad \bar A = \anull + \aone\;\;.
\ee
Expanding the F-term equations (\ref{ftermeq1})-(\ref{ftermeq2}) to linearized order in fluctuations $\fone$ and $\aone$, we obtain the equations of motion to be satisfied by the zero modes
\bea\label{zero_mode_fterm_eq1}
\bar\pa_{A^{(0)}} \aone = 0\\\label{zero_mode_fterm_eq2}
\bar\pa_{A^{(0)}} \fone + [\aone, \fnull] = 0 \;.
\eea
One can check that these are indeed the equations of motion for the fermionic zero modes in the supersymmetric Yang-Mills theory \cite{beasleyheckmanvafa1}.  Similarly, expanding the D-term equation (\ref{d_term_1}) we get the following equation for the zero modes\footnote{Technically speaking, expanding about the holomorphic and anti-holomorphic background yields a single equation of motion for the zero modes and their complex conjugates. Compatibility with the
holomorphic structure specified by the F-terms then leads to equation (\ref{dterm_zeromodes}).}
\be\label{dterm_zeromodes}
\o\wedge \pa_{A^{(0)} } \aone + \frac{i}{2} [\bar\varphi^{(0)},\fone]= 0\;.
\ee

Let us now examine the properties one can expect from a zero mode satisfying the above BPS equations.
In a local patch, and for any given zero mode solutions, we can integrate (\ref{zero_mode_fterm_eq1}) to write
$\aone=\bar\pa_{A^{(0)}} \xi$ for some regular function $\xi$ with values in the adjoint of $G_p$. Combined with the other zero mode equation, we can express any zero mode solution as
\be\label{form_of_zmodes}
\aone=\bar\pa_{A^{(0)}} \xi\quad,\quad \fone = [\fnull,\xi ] + h \ee
for some adjoint-valued $(2,0)$-form $h$ which is holomorphic with respect to $\bar\pa_{A^{(0)}}$.
Again using the local basis for $(2,0)$-forms to write
$$\fone= \fione dx \wedge dy$$ and
decomposing the zero modes in the basis $\mathfrak{g}=\mathfrak{h}\oplus_\a \mathfrak{g}_\a$ by writing
$$
\aone = \aone_\a X_\a\quad, \quad\fione=\fione_\a X_\a\quad \text{etc.},
$$
in the background discussed in (\ref{holomorphic_gauge})-(\ref{cartan_bkgnd}) the zero modes  can now be expressed in component form as
\be\label{z_modes_2}
\aone_\a = \bar \pa \xi_\a \quad,\quad \fione_\a = \a(\finull) \xi_\a + h_\a\;,
\ee
with some regular function $\xi_\a$ and a holomorphic section $h_\a$ of the canonical bundle of the surface $S$ satisfying $\bar\pa h_\a=0$.

This expression illustrates that it is possible to have zero modes with different supports on our local patch. Our primary interest will be in configurations where these zero modes localize on matter curves $\Sigma_{\a}=\{\a(\finull) = 0\}$ for positive roots $\a$.\footnote{Note that negative roots of the form $-\a$ for $\a$ positive define the same matter curve.} To define a localized chiral zero mode, let us
rewrite the above equation as
\be\label{z_modes_3}
\aone_\a = \bar \pa \Big( \frac{\fione_\a-h_\a}{\a(\finull)}  \Big) \quad,\quad \fione_{\a}\lvert_{\a(\finull) = 0} \, = h_\a\quad\text{  is holomorphic}\;.
\ee
By definition, a zero mode localized at the matter curve $\Sigma_{\a}$ is given by the above formula with $\fione_\a$ vanishing rapidly away from the curve.

The class of physically distinct zero mode wave-functions are those specified by solutions to the F-term
equations of motion, modulo gauge equivalences. For example, all zero modes $\fione_\a$ which satisfy
$\fione_\a\lvert_{\a(\finull) = 0}=h_\a$ for a given $h_\a$ with $(\fione_\a-h_\a)/\a(\finull)$
smooth everywhere on the patch are in fact gauge equivalent. To see this, consider two sets of
such zero mode wave-functions $\aone_\a,\fione_\a$ and $\bar A^{'(1)}_\a, \phi^{'(1)}_\a$. Since:
$$
\d \xi_\a = \frac{\f^{'(1)}_\a-\fione_\a}{\a(\finull)} = \frac{\f^{'(1)}_\a-h_\a}{\a(\finull)}-\frac{\phi^{(1)}_\a-h_\a}{\a(\finull)}
$$
is a function which is everywhere smooth on the patch, the difference between the two ``different'' zero modes is:
\be \label{gaugeequivs}
\bar A^{'(1)}_\a -\aone_\a = \bar \pa (\d \xi_\a)\quad,\quad \phi^{'(1)}_\a - \fione_\a = \a(\finull)\,\big( \d \xi_\a\big).
\ee
By inspection, the difference between these two solutions is an infinitesimal version of
the gauge transformation (\ref{gauge_sym}). Note that a similar argument holds for a
change of the holomorphic section $h_\a$ by
$$h_\a \to h_\a + \a(\finull) \til h_\a$$
for some holomorphic function $\til h_\a$.

Combining the above two observations, we conclude that in our local patch the space of gauge inequivalent classes of zero mode solutions localized on the curve $\Sigma_\a = \{\a(\finull) = 0\}$ is isomorphic to the space ${\cal O}_U/(\a(\finull))$, namely the space of holomorphic functions on the patch with the equivalence relation $\a(\finull) \sim 0$.\footnote{Recall that the $h_\a$'s are actually holomorphic sections of the restriction of the canonical bundle of $S$ to $U$, $K_S|_U$, a fact consistent with the above equivalence relation since $\finull$ is also a holomorphic section of $K_S|_U$. They only reduce to holomorphic functions upon local trivialization of the bundle $K_S$. Since we mostly work in a local patch in the present paper, we will sometimes abuse terminology and refer to them as ``holomorphic functions'' when the context is clear.}

\vspace{18pt}
{\it{An Example}}
\vspace{12pt}

As alluded to earlier, the gauge invariance of the physically distinct zero mode solutions
is consistent with the fact that the superpotential $W$ is independent of the K\"ahler data.
To illustrate this point, we now work in ``physical gauge'' and present an example of zero
mode solutions in a geometry where the non-holomorphic
data of the system is as simple as possible. To this end, consider a background field configuration where the
background gauge field strength is zero, and the K\"ahler metric is diagonal and flat. The corresponding values of the gauge field
$A^{(0)}$ and K\"ahler form $\omega$ can then be written in local coordinates $x$ and $y$ as:
$$
A^{(0)} = 0 \quad,\quad \o = \frac{i}{2} g \big(dx \wedge d\bar x +dy \wedge d\bar y \big).
$$
Next consider zero mode solutions localized on the matter curve $\a (\phi^{(0)}) = -x$. Plugging in the form of the localized zero
modes dictated by equation (\ref{z_modes_3}) and writing $h_\a = h_x(x,y)$, the D-term equation (\ref{dterm_zeromodes}) for the zero modes reads
$$
g \pa_x \pa_{\bar x} \big( \frac{\fione_\a - h_x }{x}\big) = \bar x \fione_\a \;.
$$
As a brief aside, although $x$ appears in the denominator of an expression
which is acted on by the Laplacian, the resulting zero mode solutions will
nevertheless be regular at the origin $x = y = 0$. This is because the
numerator $\phi_{\alpha}^{(1)} - h_{x}$ will vanish at the origin as well. As can be checked,
this equation admits solutions of the form:
$$
\fione_\a = e^{-\frac{1}{\sqrt{g}}|x|^2} h_x(y)
$$
where $h_x(y)$ is a holomorphic function of the coordinate $y$ on the curve $\{x=0\}$, on which the zero mode is localized.

The D-term constraint specifies a particular representative for the wave-function, specified
by a Gaussian falloff away from the matter curve. Comparing the profile of the matter field wave-functions
for two different values of $g$ which we refer to as $\phi_{\alpha}^{(1)}(g_1)$
and $\phi_{\alpha}^{(1)}(g_2)$, it can be shown that there exists a $\delta \xi_{\alpha}$ of the form given by equation
\eqref{gaugeequivs}. In other words, a change in the K\"ahler metric simply corresponds to a
gauge transformation of the wave-function. Moreover, the gauge inequivalent classes of
zero modes are distinguished by different holomorphic functions $h_x(y)$. In particular, we will often
find it convenient to use the basis $\{1,y,y^2,\dotsi\}$ to label the zero modes localized on the curve $\{x=0\}$.

This example also shows that the triple coupling of the zero modes localized on three matter curves is
localized at the point of intersection. To see this, first observe that the freedom to choose the
gauge, or equivalently the K\"ahler data, allows us to take the limit $g\to 0$. In this limit the
zero mode wave-functions become sharply peaked along the matter curve:
\bea\notag
\aone_\a  &=& \lim_{g\to 0} \frac{1}{\sqrt{g}} e^{-\frac{1}{\sqrt{g}}|x|^2} h_x(y) d\bar x = \pi \d(x)h_x(y) d\bar x
\\ \label{localized_delta_zero_mode}
\fione_\a &=& \lim_{g\to 0}e^{-\frac{1}{\sqrt{g}}|x|^2} h_x(y) = 2 \big ( 1-H(|x|)\,\big)  h_x(y)
\eea
where here, $\delta(x)$ is the two-dimensional delta function so that
upon writing $x = \Re(x) + \sqrt{-1} \Im(x)$, $\delta{(x)} = \delta(\Re(x))\delta(\Im(x)) = \delta{(|x|)}/|x|$. Further, $H(|x|)$ is the Heaviside step function, which satisfies:
\begin{equation}
\frac{d}{d|x|}H(|x|) = \delta{(|x|)} = \pi |x| \delta{(x)}.
\end{equation}
By an appropriate choice of K\"ahler metric, the profile of the matter field wave-functions can be made to vanish away from the matter curves. In particular, the evaluation of the Yukawa couplings is concentrated near the point of mutual overlap between the zero mode wave-functions. This localization of the matter field wave-functions near the Yukawa point is the primary justification for our approach of working in a local neighborhood containing this point. Furthermore, the $\delta$-function form of the matter field wave-functions also suggests that the triple overlap integral should be zero unless all holomorphic functions $h_\a$ are constants.


\subsubsection{The Local Cohomology Theory of Chiral Matter}
\label{The Cohomology Theory of Chiral Matter}

It turns out that there is a more formal and economical way to package the information about the equivalence classes of the localized zero modes by defining an appropriate cohomology theory. See \cite{DonagiWijnholtIII} for a related discussion.
This way of encoding the zero mode solutions helps to make the topological nature of our twisted supersymmetric Yang-Mills theory, and in particular the Yukawa coupling of the theory, manifest.

To define this cohomology, let us start by considering the total space of the canonical bundle of the K\"ahler surface $S$, $K_S \rightarrow S$ which we will refer to as
$\mathbb{K}_S$. The complex threefold $\mathbb{K}_S$ is a non--compact Calabi--Yau threefold with holomorphic $(3,0)$ form given locally as:
\begin{equation}
 \Omega= dx\wedge dy \wedge dz,
\end{equation}
where $x, y$ are holomorphic local coordinates on $S$ and $z$ is the local coordinate along the fiber.

Using the isomorphism between the canonical bundle and normal bundle:
\begin{equation}\label{isomorphism234}
 K_S \simeq N_{S/\mathbb{K}_S}
\end{equation}
we can map a $(2,0)$-form on $S$ to a $(0,1)$-form on $\mathbb{K}_S$ by
\be\label{map1}
\phi \,dx\wedge dy \mapsto \hat\k\wedge\phi
\ee
where
$$
\hat\k = g^{1\bar\imath}g^{2\bar\jmath}\: \overline{\Omega}_{\bar\imath\bar\jmath \bar z}\,d\bar z\;.
$$
In the above equation $g_{i\bar \jmath}$ is the K\"ahler metric on $S$ and we have denoted the coordinates by $x=x^1,y=x^2$.
From the fact that the holomorphic three-form is covariantly constant and that a K\"ahler metric respects the holomorphic structure, we have
$$
\bar \nabla \hat\k = \bar \pa \hat\k =\bar \pa (g^{i\bar \imath}g^{j\bar \jmath}\overline{\Omega}_{\bar\imath\bar\jmath \bar z}\,d\bar z) = 0 \;,
$$
a fact that will prove important later in order for our Dolbeault operator to correctly capture the zero mode equations. Notice that the non-compact Calabi-Yau $\mathbb{K}_S$ should really be thought of as a book-keeping device in the present construction rather than literally as the physical geometric space.

Extending the above map (\ref{map1}), we can map $(2,k-1)$-forms on $S$ to $(0,k)$-forms on $\mathbb{K}_S$ by
\be
F'\, dx\wedge dy \mapsto \hat\k\wedge F'\;,
\ee
where $F'=F'^{(0,k-1)}$ is a $K_S$-valued $(0,k-1)$-form on $S$.
In this way a pair of $(0,k)$-form and $(2,k-1)$-forms on $S$ is mapped to a $(0,k)$-form on $\mathbb{K}_S$ as
 \be
 (F,\,F' \,dx\wedge dy) \mapsto  F + \hat\k\wedge F'\;.
 \ee

By this construction, we have unified the two kinds of forms on $S$ in the standard Hodge decomposition by introducing a fictitious extra dimension. For this reason we will from now on refer to both $(0,k)$-forms and $(2,k-1)$-forms on $S$ as ``twisted $k$-forms''.

There  is also a natural way to define the integral of a twisted three-form on the patch $S$. Writing a twisted 3-form as
$$
\m =  \hat\k\wedge \hat\m
$$
where  $\hat\m$ is an $K_S$-valued $(0,2)$-form on $S$, the integral of $\m$ on the surface is simply given by
\be\label{twisted_integral}
\int_S \m \equiv \int_S \hat\m \,dx\wedge dy
\ee
as follows from the isomorphism (\ref{isomorphism234}).

Now we are ready to define the relevant cohomology. For a given background $\anull,\finull$, define the Dolbeault operator
\be
\label{D_operator}
\bar D = \bar\pa_{A^{(0)}} + \hat\k \wedge \text{adj} (\finull)
\ee
where
$\text{adj} (\finull)  = [\finull,\cdot\,]$ is the adjoint action of the background field $\finull$.

It is easy to verify the following properties of the Dolbeault operator $\bar D$: First, we observe that $\bar D$ takes twisted $k$-forms to twisted $(k+1)$-forms and that
$$
\bar D^2 = 0
$$
as a consequence of the F-term equations  (\ref{ftermeq1})-(\ref{ftermeq2}) satisfied by the background fields $\anull$ and $\finull$. Second, $\bar D$ satisfies the Leibniz rule:
\be\label{Leibnitz}
\bar D (\Psi \wedge \Psi') = \bar D \Psi \wedge \Psi' + (-1)^{\text{deg}\Psi}  \Psi \wedge \bar D \Psi' \;.
\ee
Third, we have
$$
\bar D = \bar \pa
$$
when acting on gauge singlets.

The first property shows it is meaningful to talk about a cohomology with respect to this operator. As we will see in more detail later, the last two properties ensure that the Yukawas are invariant under a change of zero mode wave-functions by a $\bar D$-exact piece, given that some mild convergence properties are satisfied by these wave-functions.

Finally we are ready to establish the connection between this cohomology and the zero mode solutions discussed in section \ref{Zero Modes Localized on Curves}. When acting on a twisted one-form
\be\label{z_mode_one_form}
\Psi^{(1)}= \aone+ \hat\k \wedge \fione\;,\ee
we have
\be
\bar D \Psi^{(1)} = \bar\pa_{A^{(0)}} \aone - \hat\k\wedge \big(  \bar\pa_{A^{(0)}} \fione + [\aone,\finull] \big)\;.
\ee
The F-term equation on the zero modes is hence equivalent to the closure of the corresponding twisted one-form under the Dolbeault operator $\bar D$.
Furthermore, what we get when acting on a scalar
\be
\bar D f = \bar \pa_{\anull} f + \hat\k \wedge [\finull,f]
\ee
is exactly an infinitesimal gauge transformation.
To sum up, the definition of equivalence classes of the zero modes can be written as
\be
\bar D \Psi^{(1)} =0 \quad,\quad  \Psi^{(1)} \sim \Psi^{(1)} +\bar D  f \quad
\ee
for any smooth adjoint-valued scalar $f$ vanishing near the boundary of our patch. From here we can conclude that the space of inequivalent zero mode solutions is isomorphic to the space of $\bar D$ cohomology $H^1_{\bar D}(U,\text{adj}({\cal P}))$, where ${\cal P}$ is the principal $G_p$-bundle.

Having specified the formal aspects of the $\bar D$ cohomology theory,
we now turn again to zero modes which are localized on the matter
curve $\Sigma_\a= \{\a(\finull) = 0\}$ of gauge enhancement.
This means we have the following expression for the localized zero modes as a twisted one-form
\be\label{z_modes_Dexpression}
\Psi^{(1)}_\a X_\a= \bar D \Big(\frac{\fione_\a-h_\a}{\a(\finull)}\, X_\a\Big) + \hat\k \wedge h_\a X_\a\quad,\quad \fione_\a\lvert_{\a(\finull) = 0} \, = h_\a,
\ee
where $\fione_\a$ vanishes rapidly off the curve $\Sigma_\a$.
Of course, this expression reduces to our old expression (\ref{z_modes_3}) when written out in components. Notice that the first term is not a pure gauge piece because the expression $(\fione_\a-h_\a)/{\a(\finull)}$ is not localized in the patch.
Furthermore, the freedom to add a $\bar D$-exact (local) piece shows that the zero modes localized on the curve $\Sigma_\a$ are indeed given by ${\cal O}_U/(\a(\finull))$, as we have seen in section \ref{Zero Modes Localized on Curves}.

The Dolbeault operator $\bar D$ also has a natural role in interpreting the D-term condition (\ref{dterm_zeromodes}). Using the metric
$$
ds^2= g_{i\bar\jmath}\,dx^i\,dx^{\bar\jmath} + \text{det}(g)^{-1} dz\,d\bar z
$$
on the threefold $\mathbb{K}_S$, we see that the D-term equation (\ref{dterm_zeromodes}) is nothing but the statement that $\bar D^\dag \Psi^{(1)}=0$. Hence the physical wave-function is really the harmonic representative of the $\bar D$-cohomology. Nevertheless, as we have argued earlier and as we will see explicitly later, the topological nature of the Yukawa coupling in our theory guarantees that the answer does not depend on which representative, harmonic or not, we pick in the cohomology for the zero mode wave-function.

\subsection{The Yukawa Coupling}
\label{The Yukawa Coupling}
Finally we are ready to compute the Yukawa coupling of interest. Putting (\ref{separate_bkgnd}) into the superpotential (\ref{superpotential}) and using the BPS equations satisfied by the background and the zero modes, we get the following expression for the superpotential
\be\label{yuk_1}
W_{\text{Yuk}} = \int_U \Tr ( \aone\wedge \aone \wedge\fone)\;.
\ee
In the language of the twisted one-forms of section \ref{The Cohomology Theory of Chiral Matter}, using equations (\ref{twisted_integral}) and (\ref{z_mode_one_form}) the above equation can be written in a more symmetric-looking form as:
\be
W_{\text{Yuk}} = \int_U \Tr (\Psi^{(1)}\wedge\Psi^{(1)}\wedge\Psi^{(1)})\;.
\ee

As explained in the last subsection, the gauge invariance of the superpotential ensures that $W_{\text{Yuk}}$ is well-defined in the sense that  it only depends on the $\bar D$-cohomology of the wave-function $\Psi^{(1)}$.
Indeed, the chiral superfields localized on the matter curve $\Sigma$ have the cohomological interpretation as elements of the group $\text{Ext}^1({\cal E},{\cal E}')$ where ${\cal E}$, ${\cal E}'$ are the coherent sheaves associated with the holomorphic gauge bundles along the seven-branes $S$ and $S'$ which intersect on the curve $\Sigma$. The Yukawa coupling is in this case given by the Yoneda pairing of the Ext-groups:
$$
\text{Ext}^1({\cal E},{\cal E}')\otimes\text{Ext}^1({\cal E}',{\cal E}'')\otimes\text{Ext}^1({\cal E}'',{\cal E})\to\C\;.
$$
See for example \cite{altman} for a discussion of the Yoneda pairing, and \cite{Sharpe:2003dr} for a discussion in the context of superpotentials for D-branes. In the context of F-theory GUTs, a formulation of Yukawas in terms of Yoneda pairings has been discussed for example in \cite{DonagiWijnholtIII,Tatar:2009jk}.

Let us now turn to the physical meaning of this triple coupling. While the given zero modes satisfy the linearized equation of motion (and hence the absence of mass terms), there is no guarantee that the corresponding deformation remains unobstructed beyond the linearized level. In fact, a non-zero value for this triple coupling means the deformation is obstructed. The Yoneda pairing formulation also demonstrates that the gauge bundle data is already taken into account in computing the form of the Yukawa couplings.

This last point is somewhat at odds with the expectation of \cite{Heckman:2008qa} reinforced in \cite{Font:2009gq} that because the gauge field fluxes distort the profile of the matter wave-functions, the Yukawa couplings should also be distorted. Indeed, much of this distortion can be ascribed to the D-term equations of motion for the matter field wave-functions. Since the contribution to the Yukawa coupling are independent of non-holomorphic data, it follows that only the F-term equations of motion (\ref{zero_mode_fterm_eq1})-(\ref{zero_mode_fterm_eq2}) are relevant in discussions of the Yukawa coupling. On the other hand, in a local patch, compatibility of the complex structure of the surface $S$ and the gauge bundle through the equation $\bar\pa_{A^{(0)}}^2=0$ ensures that we can pass to a holomorphic gauge in which $\anull=0$. In particular, this suggests that it is always possible to recast our computation in terms of a configuration where the local profile of the fluxes are absent.

Indeed, in the remainder of this section we will show that gauge field fluxes alone do not alter the form of the Yukawa couplings. To reach this conclusion, we first study two representative examples in ``physical gauge'' where we solve for wave-functions which satisfy both the F- and D-term equations of motion. In physical gauge, the absence of a change in the structure of the Yukawas can be ascribed to a non-trivial cancellation between various contributions to the Yukawa overlap integrals. To provide further evidence for this general point, we next present a residue formula which makes the quasi-topological nature of the Yukawa coupling manifest.

\subsubsection{Two Examples}
\label{Two Examples}

We now present two examples of Yukawa couplings between zero modes, first in the absence of a background gauge field flux, and then in the presence of a gauge field flux with non-trivial first order gradients. To be concrete, we work
in physical gauge by specifying zero mode wave-functions which satisfy both the F- and D-term equations of motion.

For simplicity, we consider a geometry with a single point of $G_p=SU(3)$ enhancement and choose a flat and diagonal K\"ahler form:
\be
\o = \frac{i}{2} g \big(dx \wedge d\bar x +dy \wedge d\bar y \big) \;.
\ee
We consider a background value for $\varphi^{(0)}$ such that $G_p = SU(3)$ is broken down to its maximal torus $U(1)^2$ at generic points of the patch through the vev:
\be\label{su3_bkgrnd}
\fnull = \finull dx\wedge dy = \frac{1}{3}\bem -2x + y &0&0\\ 0&x+y&0\\ 0&0&x-2y\eem dx\wedge dy.
\ee
Indeed, from the F-term equation (\ref{ftermeq2}), we see that $\finull \in \mathfrak{h}$ forces the
gauge background $\anull$ to be in the Cartan subalgebra at generic points as well. To be entirely explicit we work in the gauge in which the gauge field is real.

To set notation, let $\a_1$ and $\a_2$ denote the two roots of $SU(3)$ such that $X_{\a_1}$ and $X_{\a_2}$ are given by $3 \times 3$ matrices with a single non-zero entry respectively
in the $1$-$2$ and $2$-$3$ entry. In this notation, the relevant zero modes are labelled by the $1$-$2$, $2$-$3$ and $3$-$1$ entries of a $3 \times 3$ matrix which respectively correspond to the roots $\a_1$, $\a_2$ and $\a_3 \equiv -\a_1-\a_2$.

We will solve for the zero mode wave-functions in the following way: From the expression for the localized zero modes (\ref{z_modes_3}), or equivalently (\ref{z_modes_Dexpression}), we see that the zero modes satisfy
\be
\Psi^{(1)}_\a X_\a= \bar D \Big(\,\frac{\fione_\a}{\a(\finull)} X_\a\Big) \quad,\quad \text{on}\;\;\;U\setminus\Sigma_\a
\ee
everywhere away from the matter curve $\Sigma_\a$. Then the D-term equation implies
\be\label{dterm_in_example}
\bar D^\dag \bar D \Big(\frac{\fione_\a}{\finull_\a} X_\a\Big) = 0  \quad,\quad \text{on}\;\;\;U \setminus\Sigma_\a \;.
\ee
The `Laplacian' operator $\bar D^\dag \bar D$ is positive definite, and with our Ansatz for the K\"ahler form it reads
$$
(\bar D^\dag \bar D)_\a = -g \Big( (\pa_x + i A_{\a,x})(\pa_{\bar x} + i A_{\a,\bar x})+(\pa_y + i A_{\a,y})(\pa_{\bar y} + i A_{\a,\bar y})\Big) + |\finull_\a|^2
$$
when acting on the $X_\a$ component of the zero mode. The reason that the equation (\ref{dterm_in_example}) can have  a  solution is that it does not hold on the matter curve. Using this equation, we next specify the behavior of the zero mode on the matter curve through the relation $h_\a=\phi^{(1)}_\a\lvert_{\a(\finull) = 0}$. This then determines the full profile of the zero mode solution.

\vspace{18pt}
{\it{First Example: Without Fluxes}}
\vspace{12pt}

Having discussed the general set-up, we now compute the Yukawa couplings in a configuration with all gauge field fluxes turned off. In particular, we choose the gauge
$$
A^{(0)}_{\a_1}=A^{(0)}_{\a_2}=0\;.
$$
In this case the zero mode solutions are of the Gaussian form we have seen earlier in (\ref{localized_delta_zero_mode}). Explicitly, they are
\bea\notag
\aone_{\a_1} &=& \frac{1}{\sqrt{g}} e^{-\frac{1}{\sqrt{g}}|x|^2}  y^n d\bar x\quad,\quad \fone_{\a_1} =e^{-\frac{1}{\sqrt{g}}|x|^2}  y^n dx\wedge dy \\ \label{commutative_simple_wavefunction}
\aone_{\a_2} &=& -\frac{1}{\sqrt{g}} e^{-\frac{1}{\sqrt{g}}|y|^2}  x^m d\bar y\quad,\quad \fone_{\a_2} =e^{-\frac{1}{\sqrt{g}}|y|^2} x^m dx\wedge dy \\ \notag
\aone_{\a_3} &=& -\frac{1}{\sqrt{2g}} e^{-\frac{1}{\sqrt{2g}}|x-y|^2}   (x+y)^\ell (d\bar x-d\bar y) \quad,\quad \fone_{\a_3} =e^{-\frac{1}{\sqrt{2g}}|x-y|^2}  (x+y)^\ell dx\wedge dy
\eea

The Yukawa coupling (\ref{yuk_1}) in this case is:
 \bea\notag
W_{\text{Yuk}} &=& \int_S \Tr ( \aone\wedge \aone \wedge\fone)\\\notag
 & =& \frac{1+\sqrt{2}}{g} \int  \,e^{-\frac{1}{\sqrt{g}}(|x|^2+|y|^2+|x-y|^2/\sqrt{2})}  y^n\, x^m\,(x+y)^\ell  dx\wedge d\bar x \wedge dy\wedge d\bar y
 \eea

It follows from this expression that the integrand admits a natural rephasing symmetry under the $U(1)$ action
$$
x\to e^{i\th} x\quad,\quad y\to e^{i\th} y.
$$
In particular, this implies that the Yukawa coupling vanishes between all zero modes, except when $\ell=m=n=0$. Further note that the answer is indeed independent of the K\"ahler data $g$ as we expected. In particular we are free to take the zero-volume limit $g\to 0$ in which the wave-functions converge to $\delta$-functions (\ref{localized_delta_zero_mode}). An explicit evaluation of the integral yields
\be\label{no_fluxes}
\frac{1}{(2\p )^2} \,W^{\ell mn}_{\text{Yuk}} = \begin{cases} 1 & \ell=m=n=0\\ 0 &\text{otherwise}
\end{cases}\;.
\ee

\vspace{18pt}
{\it{Second Example: Non-Constant Fluxes}}
\vspace{12pt}

We now study the effects (or lack thereof) of gauge field fluxes on Yukawa couplings. With $\finull$ as before, we next consider a background gauge field which takes values in the Cartan subalgebra of the Lie algebra of $G_p = SU(3)$. The projection of this direction in the Cartan onto the roots $\a_1$ and $\a_2$ of $\mathfrak{su}(3)$ are specified as:
$$
\a_1(A^{(0)})= \a_2(A^{(0)}) = i \e_1 (\bar y^2 d x-\bar x^2 d y + \e_2(\bar y^3 d x+\bar x^3 d y )\, ) + \text{c.c.}
$$
which describes a background gauge field strength of the form:
$$
\a_1(F^{(0)}) = \a_2(F^{(0)}) =  i\e_1\Big( 2(\bar x dy\wedge d\bar x-\bar ydx\wedge d\bar y) -3\e_2(\bar x^2 dy\wedge d\bar x+\bar y^2 dx\wedge d\bar y) \Big)+\text{c.c}\;.
$$
Note in particular that this gauge field strength has non-trivial gradients, and so will produce three tensor index objects of the type required
by the flavor hierarchy argument of \cite{Heckman:2008qa}. As we now show, however, such fluxes do not alter the structure of the Yukawa couplings.

We begin by solving for the zero modes in this background. The solution to the D-term equation (\ref{dterm_in_example}) is to first order in the expansion of small flux given by:
\bea\notag
\fione_{\a_2}(x,\bar x,y,\bar y)&=&e^{-\frac{1}{\sqrt{g}}|y|^2 } \Big\{h_{\a_2}(x) + \e_1 \Big[ \big(-\bar x^2y+\e_2 \bar x^3y \big)  h_{\a_2}(x) \\
\notag&&+\big( -2  \bar xy +3 \e_2  \bar x^2y \big) \sqrt{g} h_{\a_2}'(x)+\big(-2 y +6   \e_2\bar xy \big) g\,h_{\a_2}''(x) +
6\e_2 y  g^{3/2} h'''_{\a_2}(x)\Big] \\\notag
&&+\bar \e_1\, \Big[\big( x^2 \bar y  -\bar\e_2 x^3 \bar y  \big) h_{\a_2}(x)+\big( \frac{1}{2} y^2
+\frac{1}{3}\bar \e_2y^3  \big) \sqrt{g} h_{\a_2}'(x)\Big] + {\cal O}(F^2)\Big\}\;.
\eea
The symmetry of the fluxes we chose implies that, for the given holomorphic function $h_{\a_1}(y)$, the zero mode $\fione_{\a_1}(x,\bar x,y,\bar y)$ is obtained by exchanging $x\leftrightarrow - y$ in the above formula. To simplify our analysis, we now restrict to the Yukawas in the special case where $\fione_{\a_3}\lvert_{\a_3(\finull) = 0}\, =1 $. This is an appropriate description for models with a single MSSM Higgs field localized on the curve $\{\a_3(\finull) = 0\}$, with some number of generations on the other matter curves. The corresponding profile for this zero mode is then given by:
\bea\notag
\fione_{\a_3}(x,\bar x,y,\bar y)&=&e^{-\frac{1}{\sqrt{2g}} |x-y|^2}\Big\{1+\text{$\e_1 $} \Big[ -\frac{1}{6 }(x-y) (\bar x-\bar y)^2-\frac{1}{2}(x-y) (\bar x+\bar y)^2
\\ \notag&&+\, \e_2|x-y|^2\,\big(\,\frac{1}{16} (\bar x-\bar y)^2+\frac{3}{8} (\bar x+\bar y)^2\big)  \Big]\\ \notag&&
+\,\bar \e_1\Big[\,\frac{2}{3} \sqrt{2g} \,(x-y)+\frac{(x-y)^2 (\bar x-\bar y)}{6}+\frac{(x+y)^2 (\bar x-\bar y)}{2 }
\\ \notag
&&-\, \bar \e_2 \big(\,\frac{3 \sqrt{2g} \,(x-y)^2}{16}+\frac{1}{16} (x-y)^3 (\bar x-\bar y)\\ \notag
&&+\, \frac{3}{8} (x-y) (x+y)^2 (\bar x-\bar y)-\frac{1}{4} (x+y)^3 (\bar x+\bar y) \big) \Big]+{\cal O}(F^2)\Big\}\;,
\eea
and the $\aone$ wave-functions are given by the above expression and the zero mode formula (\ref{z_modes_3}). To reliably estimate the $2$-$2$ entry of the Yukawa matrix, it is actually necessary to include second order corrections from the fluxes. These expressions are not particularly illuminating, and so for the sake of brevity we shall only present the explicit expressions for the first order corrections
to the wave-functions.

From the form of the zero mode wave-functions, it follows that there is an effective $U(1)$ rotational symmetry which is broken by the spurion
parameters $\e_1$ and $\e_2$. Indeed, under the diagonal $U(1)$ which acts as
$$
x\to e^{i\th} x\quad,\quad y\to e^{i\th} y\;,
$$
the flux parameters $\e_1$ and $\e_2$ both carry charge one. This is in accord with the analysis of \cite{Heckman:2008qa} that fluxes can generate an effective Froggatt-Nielsen mechanism. See also \cite{Randall:2009dw} for further discussion on a discussion of the associated effective field theory.

Although consistent with this selection rule, in the actual computation of wave-function overlaps, we find that the resulting Yukawa structure remains unchanged by the background gauge field flux. To illustrate this point, we consider a two generation model in which the wave-functions on the matter curves $x = 0$ and $y = 0$ are specified by the associated $h_{\a}$ as:
$$
h_{\a_1}(y) = y^{m}\quad,\quad h_{\a_2}(x) = x^{n}
$$
for $m,n = 0,1$, while that of the ``Higgs'' wave-function is specified as:
$$
h_{\a_3} = 1.
$$
In this case, we find that the resulting Yukawa is given by:
\be\label{commutative_fluxes_rank1}
\frac{1}{(2\p )^2} \,W^{mn}_{\text{Yuk}} = \begin{cases} 1 & m=n=0\\ 0 &\text{otherwise}
\end{cases}\;.
\ee
In other words, comparing this result with that of equation \eqref{no_fluxes}, we find that there is no difference in the two
Yukawa structures. This result is due to a number of non-trivial cancellations between contributions to the overlap integral, and
in the remainder of this section, our aim will be to make such cancellations more transparent.

\subsubsection{Yukawas From Residues}
\label{Yukawa Coupling as Residue}

The examples presented in the previous subsection strongly suggest that gauge field fluxes do not distort the structure of the Yukawa couplings. In this subsection we generalize this result by showing how Yukawa couplings can be formulated in terms of a residue formula defined in a neighborhood containing the Yukawa enhancement point $G_p$. This quasi-topological formulation of Yukawa couplings will allow us to prove a more general ``Rank Theorem'' that when only gauge field fluxes are activated, the rank of the Yukawa matrix is completely determined by the intersection theory of the matter curves, and so in particular, is independent of the gauge field flux.

To obtain a useful formula for this triple coupling, let us focus on the point of intersection of the two matter curves $\a_1(\finull) = 0 $ and $\a_2(\finull) = 0 $. Here we assume that these two curves intersect transversely. Notice that this necessarily implies the presence of the third matter curve $\a_3(\finull) = 0 $, as long as $\a_3=-\a_1-\a_2$ is in the root system. Our starting point for the Yukawa coupling is the overlap integral for three twisted one-forms:
\bea\notag
W_{\text{Yuk}} &=& \int_U \Tr(\Psi^{(1)}\wedge \Psi^{(1)}\wedge \Psi^{(1)}) \\ \notag
&=& \int_U \Psi_{\a_1}\wedge \Psi_{\a_2}\wedge \bar D  \Big(\frac{\fione_{\a_3}-h_{\a_3}}{\a_3(\finull)}\Big)
+\int_U  h_{\a_3}\, \Psi_{\a_1}\wedge \Psi_{\a_2}\wedge \, dx\wedge dy\\\label{pre_residue}
&=&  \int_U \; h_{\a_3}\, \bar \pa_{A^{(0)}}  \Big(\frac{\fione_{\a_1}-h_{\a_1}}{\a_1(\finull)}\Big)\wedge \bar  \pa_{A^{(0)}}  \Big(\frac{\fione_{\a_2}-h_{\a_2}}{\a_2(\finull)}\Big) \, dx\wedge dy
\eea
where in the above formula we have dropped the first term in the second line
by first rewriting it as a boundary term and then using the property that $ \Psi_{\a_1}\wedge \Psi_{\a_2}$ is localized around the intersection point of the two matter curves $\Sigma_{\alpha_1}$ and $\Sigma_{\alpha_2}$.

Let us now focus on the remaining term in the integral. From the form it is rather clear that it can be rewritten a residue integral. See Appendix \ref{Calculation of the Residue Formula} for the actual calculation. The final answer reads
\be\label{residue_formula_commutative}
\frac{1}{(2\p i)^2}W_{\text{Yuk}}  =  \text{Res} \Bigg(\frac{ h_{\a_1}h_{\a_2}h_{\a_3} }{\a_1(\finull) \a_2(\finull)}\Bigg)\;\;.
\ee

This formula makes immediately obvious the two  major conclusions drawn from our previous explicit calculations of lines (\ref{no_fluxes}) and (\ref{commutative_fluxes_rank1}). The first is that although the actual wave-functions given by $\fione_\a$ are different, the Yukawa coupling has to be the same since it only depends on the holomorphic section $h_\a = \fione_\a\lvert_{\a(\finull) = 0}$ of the canonical bundle, namely the wave-function evaluated on the matter curve. Second, as long as our matter curves are ``simple'' and hence can be  described by $x=0$ and $y=0$ locally near the intersection point, the only set of three zero modes having non-vanishing Yukawa coupling is just
$$
h_{\a_1}= h_{\a_2}=h_{\a_3} =1\;.
$$

\subsubsection{A Rank Theorem}
\label{A Rank Theorem}

As we mentioned at the end of section \ref{Yukawa Coupling as Residue}, the residue expression for the Yukawa coupling (\ref{residue_formula_commutative}) makes manifest that the rank of the Yukawa matrices in the examples we computed earlier is one. Now we will sharpen this statement in the following way.

{\it In the theory described by the superpotential (\ref{superpotential}), if the matter curves are smooth, reduced, and intersect transversely at point $p$, then the rank of the Yukawa coupling associated with the intersection point $p$ is at most one. }

To simplify the formal analysis, let us fix the zero mode on the third matter curve to be $h_{\a_3} =1$, motivated by the phenomenological application to the MSSM and the fact that there is only one Higgs mode present in a given class of Yukawas. In other words, we associate the curve $\Sigma_{\a_3}$ to the Higgs and the other two curves  $\Sigma_{\a_1}$, $\Sigma_{\a_2}$ to chiral generations.

Let $h_{\alpha_1,m}$, $h_{\alpha_2,n}$ ($m,n=1,2,\dots$)
be the holomorphic sections corresponding to the various generations. From the residue formula (\ref{residue_formula_commutative}) we see that the Yukawa matrix is proportional to
\begin{equation}\label{eq:matrixY}
 W_{\text{Yuk}}^{mn}=\mathrm{Res}\Big(\frac{h_{\a_1,m} h_{\a_2,n} }{\a_1(\finull)\a_2(\finull)}\Big)
\end{equation}
The {local duality theorem} (see ref.\,\cite{griffithsharris} pages 659 and 693) states that the bi-linear pairing defined by the above residue
\begin{equation*}
	\langle f,\, g\rangle = \mathrm{Res}_{p}\Big( \frac{f\, g}{\a_1(\finull)\a_2(\finull)}\Big)
\end{equation*}
 is {a perfect pairing} in the ring
\begin{equation}
 \mathcal{O}/\mathcal{J}\equiv \mathcal{O}/(\a_1(\finull),\a_2(\finull))\;.
\end{equation}
In particular, all localized modes $h\in \mathcal{J}$  decouple from the cubic Yukawa interaction, again a fact reflecting the gauge invariance of the theory.
We therefore obtain:
\begin{equation}\label{eq:rank4}
\mathrm{rank}\langle \cdot, \cdot\rangle=\dim_\mathbb{C} \mathcal{O}/(\a_1(\finull),\a_2(\finull)) \equiv (\Sigma_{\alpha_1}\cdot\Sigma_{\alpha_2})_p
\end{equation}
where $(\Sigma_{\alpha_1}\cdot\Sigma_{\alpha_2})_p$ stands for the intersection number of the curves (divisors)
$\Sigma_{\alpha_1}$, $\Sigma_{\alpha_2}$ at the point $p$. Indeed, the last equality in equation \eqref{eq:rank4} is precisely the definition of the intersection number of two distinct curves on a complex surface.
 Comparing equations \eqref{eq:matrixY} and \eqref{eq:rank4}, we get the following result
\smallskip

\textbf{Rank theorem.} {In the theory with superpotential given by (\ref{superpotential}), the rank of the Yukawa matrix $W_{\text{Yuk}}^{mn}\,$ associated with a given intersection point $p$ is equal to or less than the intersection degree
$(\Sigma_{\alpha_1}\cdot\Sigma_{\alpha_2})_p$ of the matter curves at $p$.}
\smallskip

\section{$H$-Flux and Non-Commutative F-Terms} \label{sec:FTOP}

In the previous section we studied the Yukawa coupling of the seven-brane field theory with background gauge fluxes turned on, and concluded that the Yukawa coupling does not change in the presence of the gauge field fluxes. In a certain sense, however, this
analysis is incomplete because there are more general fluxes present in generic F-theory compactifications.
In this section, we focus on the three-form fluxes $H=H_R+\tau H_{NS}$ in F-theory and ask the question of how their presence deforms the superpotential of the seven-brane gauge theory. Indeed, from the tadpole condition of F-theory compactified on a Calabi-Yau fourfold $Z$  \cite{Sethi:1996es}
$$
\int_X H_{R} \wedge H_{NS} + N_{D3} = \frac{\chi(Z)}{24}\;,
$$
where $X$ denotes the base of the fourfold,
$N_{D3}$ is the number of $D3$-branes and $\chi(Z)$ is the Euler characteristic of the fourfold, we see that the F-theory three-form fluxes will generically be present.

The $(1,2)$ component of the $H$-flux which we denote by $H = H_R + \tau H_{NS}$ is the object which enters the four-dimensional $\mathcal{N} = 1$ superpotential as a chiral superfield. For example, in compactifications of type IIB string theory, $H$ enters the closed string superpotential as \cite{Gukov:1999ya}:
\begin{equation}
W=\int \Omega \wedge H
\end{equation}
where $\Omega$ is the holomorphic three-form on a Calabi-Yau threefold. This expression generalizes to F-theory, and so it is appropriate to treat $H$ in the same way there as well. From this perspective, it is therefore quite natural to expect $H$ to enter into the superpotential for the open string sector as well.

By analyzing the superpotential of probe branes in the presence of background fluxes,
we will find that in the presence of $H$-fluxes, the F-term equations of motion are modified.
This induces a non-commutative deformation of the topological theory so that for
holomorphic coordinates $X^{i}$, we have:
$$
[X^{i},X^{j}] = \theta^{ij}.
$$
Here, the field theory parameter $\theta^{ij}$  parameterizing the non-commutative deformation satisfies the relation:
\be\label{poisson_H_flux}
\nabla_{i }\theta^{jk}=H_i^{\;\;jk}\;,
\ee
to leading order in $H$. In the above,
$$
H_i^{\;\;\,jk}=g^{j{\bar\jmath }}g^{k \bar{k}} H_{i{\bar\jmath}\bar{k}}.
$$

Our strategy for establishing this result is as follows. Since we are mainly interested in local
questions, we take the ``internal directions'' of the compactification to be $\mathbb{C}^3$ and ignore the effects of monodromy on $\tau_{IIB}$
by an $SL(2,\mathbb{Z})$ transformation for simplicity. Indeed the
relation (\ref{poisson_H_flux}) would suggest that $\theta$ should
be a section of a bundle.  To further simplify the discussion
we take $\tau=\t_{IIB}$ to be constant.  We first show that a
non-commutative deformation of the topological theory is generated
for non-zero $H_R$ as a consequence of the Myers effect \cite{Myers:1999ps}.
Furthermore since only chiral fields can appear in the topological B-model
sector, and since $H$ is a chiral field, this will show that the relation
also holds when $\tau H_{NS}$ is added to $H_R$.  Next, we interpret
the presence of $H$-flux in terms of the topological B-model. As a final check on this relation, we provide an independent argument for this relation by studying realizations of the Poisson bi-vector $\theta$ in the context of generalized complex geometry. In the next section we will study how this non-commutative deformation affects the field theory and in particular its Yukawa couplings.

\subsection{$H_R$-Flux and the Myers Effect}
\label{R_flux and the Myers Effect}
In this section we will show how $H_R$-flux leads to superpotential
terms for D3-branes. Since the D3-brane corresponds to a probe of the topological B-model, this will allow us to deduce the presence of a non-commutative deformation of the seven-brane F-terms. Indeed, as we show in the next subsection, the D3-branes can also be described by defects in the non-commutative deformation of holomorphic Chern-Simons theory.

First let us briefly recall the Myers effect \cite{Myers:1999ps}.  Consider
type IIA string theory with four-form G-fluxes turned on.  A stack of $N$ D0-branes feels
a potential due to this G-flux given to leading order in constant weak $G$ by
$$V_{D0}= \Tr (X^I X^J X^K)G_{0IJK}+...$$
where $0$ denotes the time direction and $X^{I}$ denotes a $U(N)$ valued field of the D0-brane theory whose eigenvalues indicate the positions of the D0-branes in the nine spatial directions.

This configuration admits a T-dual description. Focussing on fluxes with spatial indices in $\mathbb{C}^{3}$, if we compactify the three remaining spatial directions on a $T^3$, applying three T-dualities to the D0-brane configuration yields a D3-brane probe in type IIB in the presence of $H_R$ flux (given
by the Hodge dual of the seven-form field strength) with potential:
$$V_{D3}= \Tr(X^IX^J X^K) \epsilon_{IJK}^{\text{ \ \ \ \ \ }LMN} H_{LMN}+...$$
where $I,...,N = 1,...,6$ are indices labelling the six transverse directions to the stack of D3-branes.

As we have  explained in section \ref{F-theory Yukawas and Seven-Branes}, $H_R$ is a (1,2) form  in the case of interest to us.
In terms of the local complex coordinates $X^{i}$
in the normal direction to the D3-brane (viewed as the internal geometry of type IIB string), we can thus write the potential as
$$V_{D3}= \Tr(X^iX^{\bar j}X^{\bar k}){H_{i \bar j\bar k}}+...$$
Moreover, in the case of interest to us, ${\cal N}=1$ supersymmetry is preserved. Assuming a canonical K\"ahler potential, the potential for the stack of probe D3-branes can then be written in terms of a superpotential as:
$$V_{D3}=\sum (\partial_iW)(\partial_{\bar i}\overline{W}).$$
For a stack of $N$ D3-branes, this implies that the leading
terms of the superpotential in the $H$ expansion are
$$\epsilon^{ijk}\partial_kW=[X^i,X^j]-H^{\;\;ij}_k X^k+...\;,$$
which indicates that the F-term moduli space for the stack of D3-branes is non-commutative. Indeed, the F-term
equation of motion $[X^i,X^j]=0$ for a probe of commutative space is now replaced by:
\begin{equation} \label{NCd3}
[X^i,X^j]=\theta^{ij}=H^{\;\;ij}_kX^k+...\;.
\end{equation}
This shows that to leading order in weak $H$ we have:
$$\nabla_k \theta^{ij}=H^{\;\;ij}_k$$
In a local patch, this can be solved by
$$\theta^{ij}=g^{i\bar{i}} g^{j\bar{j}}B_{\bar{i} \bar{j}}+const.\;,$$
where $B$ is the RR $(0,2)$-form gauge potential.
The constant can be absorbed into a gauge choice
for $B$ and we will choose it such that $B$ vanishes when $H$ vanishes.

This non-commutativity is reminiscent of what is found in the D-brane sector of the superstring in the presence of $B_{NS}$
\cite{Connes:1997cr,seibergwitten}. However, there are differences from the present case. First, we have non-commutativity even when
we only have $B_{R}$. Secondly, the non-commutativity is only in the topological sector, rather than the full
superstring theory. 

Even though this formula was derived for $H_R \neq 0$ and $H_{NS} = 0$, since $W$ is holomorphic with respect to
$H = H_R + \tau H_{NS}$, it follows that the non-commutative deformation of the
F-term equation of motion for the D3-brane theory extends to more general $H$ (which
is consistent with our notation above, where we occasionally omitted the subscript from $H$).

\subsection{$H$-Flux and the B-Model}

Having motivated equation (\ref{poisson_H_flux}), we now explain why this means that the topological B-model
which computes such superpotential terms lives in non-commutative space.
In \cite{kapustin,deformation} the deformation of the topological B-model by turning on an element of $H^0(\Lambda ^2 T^{1,0})$ is considered.  From the viewpoint of the Kodaira-Spencer theory \cite{bcov}, these correspond to giving vevs to the ghost fields (more generally one could
consider deformations given by $H^p(\wedge ^qT^{1,0})$), and a deformation of the complex structure by
$$
\bar \pa \to \bar \pa ' = \pa_{\bar\jmath}+\th ^{ij}\pa_j\;.
$$
More explicitly, in local coordinates the deformed $\bar\pa'$ acts on $\oplus_{m,n} H^m(\L^n T^{1,0})$ as
$$
\bar \pa ' : \O^{i_1 \dotsi i_n}_{\;\;\;\;\;\;\;\;\;\;\bar\jmath_1\dotsi\bar\jmath_m}  \mapsto \pa_{\bar\jmath} \O^{i_1 \dotsi i_n}_{\;\;\;\;\;\;\;\;\;\;\bar\jmath_1\dotsi\bar\jmath_m}   + \th ^{ij}\pa_j   \O^{i_1 \dotsi i_n}_{\;\;\;\;\;\;\;\;\;\;\bar\jmath_1\dotsi\bar\jmath_m}
$$
with all uncontracted indices anti-symmetrized. The integrability condition $\bar\pa'^2 = 0$ can be written as two conditions:
$$
\bar\pa \th  = 0 \quad,\quad \th ^{i\ell}\pa_\ell \th ^{jk} + (\text{anti-symmetrization of } ijk)  = 0 \;.
$$
This shows that $\th ^{ij}$ has to be a holomorphic Poisson bi-vector.

As has been shown in \cite{deformation}, this deformation of the topological string makes
all the open string sectors non-commutative. In other words, we obtain the non-commutative
deformation of the holomorphic
Chern-Simons theory living on the six internal directions of D9-branes wrapping a Calabi-Yau manifold.
More precisely, the three complex dimensions now have non-commutative coordinates satisfying
$$[X^i,X^j]=\theta^{ij}.$$
Similarly, by turning on fluxes on the D9-brane we obtain its reduction
to D7-, D5- and D3-branes.

There is yet another way to see this. As was discussed in  \cite{Gukov:1999ya}, turning on fluxes gives rise to a superpotential term
$$W=\int \Omega \wedge H\;,$$
where $\Omega$ is the background value of the KS field  $A'$.  Giving a vev
to $\theta^{ij}$ corresponds to a deformation of the form $\Omega\rightarrow \Omega +\theta'$
where $\theta'$ is a one-form given by
$$\theta'_k= \theta^{ij}\Omega_{ijk}.$$
Moreover, the totality of RR fields unify to a sum over all odd-dimensional field strengths.
It is therefore natural to generalize the above superpotential to all such fields. Flux
induced superpotentials for D-branes have been discussed in \cite{martucci}. Given the
above deformation of $\Omega$ by $\theta'$ we get an additional contribution from
the five-form field strength of the D3-brane gauge potential.  This implies that
if we have a stack of D3-branes and move it from one internal point to another, we get an additional
contribution to the superpotential (in addition to the usual contribution from
the $\mathcal{N} = 4$ theory, $\epsilon_{ijk} \Tr( [X^i,X^j]X^k)\;$) given by
$$\Delta W = W(X)-W(X_0)=- \int_{X_0}^X \theta'$$
In other words, we have
$$\frac{\partial W}{\partial X^k}=([X^i,X^j]-\theta^{ij})\Omega_{ijk}\;,$$
which agrees with what we have obtained by identifying the topological string deformation parameter $\theta$
with the $B$ field. See \cite{martucci} for additional discussion on flux induced superpotentials for D-branes.

Having analyzed the behavior of a stack of probe D3-branes in the presence of a background $H$-flux, let us now
briefly mention supersymmetric configurations for other probe branes. It
would be interesting to analyze whether non-commutative deformations could be used to engineer novel examples of
supersymmetry breaking. Returning to equation \eqref{NCd3}, in the case of a single D3-brane, the
usual commutator of the $\mathcal{N} = 4$ theory vanishes, and we are left with the condition $\theta^{ij}(X) = 0$. In the case of a five-brane which
wraps a one complex dimensional submanifold of the internal geometry, the worldvolume coordinate
clearly commutes with itself. The two complex coordinates in directions normal to the five-brane
will not in general commute, and so just as for the D3-brane theory, a supersymmetric configuration
corresponds to the special case where it wraps a locus along which $\theta$ vanishes. For
example, if the five-brane wraps a complex curve defined by the intersection of the two
divisors $\{f_1 = 0\}$ and $\{f_2 = 0\}$, then as shown in \cite{kapustin}, we have the condition $\theta(d f_1 , d f_2) = 0$, where here we are using the fact that the holomorphic bi-vector acts naturally on pairs of holomorphic one-forms. Next consider the case of seven-branes. In this case, there is only a single direction normal to the seven-brane, so the condition that it commutes with itself is trivially satisfied. On the other hand, the internal worldvolume of the seven-brane is two complex-dimensional, and so will now be non-commutative. The theory on a nine-brane
is similar but of less interest for F-theory. Since the topological B-model captures F-terms of the seven-brane theory, we conclude that the effects of the $H$-fluxes on F-terms such
as the Yukawa couplings are computed by a non-commutative deformation of the partially twisted seven-brane theory.

\subsection{Poisson Bi-Vectors and Generalized K\"ahler Geometry}

In the previous subsections we have argued that in the weak field limit, the presence of a background $H$-flux induces a non-commutative deformation of the seven-brane F-terms. To further motivate equation (\ref{poisson_H_flux}), we now derive the same relation between the Poisson bi-vector $\theta$ and $H$-flux in
the context of generalized K\"ahler geometry.

We begin by first reviewing a few facts about generalized K\"ahler geometry. The manifold with data $(I_\pm, g, H)$ was first introduced by the authors of \cite{Gates:1984nk} as the target manifold of ${\cal {N}}=(2,2)$ sigma-models with worldsheet coupling to a B-field such that $dB=H$ locally. The two complex structures $I_\pm$ and the bi-hermitian metric $g$ satisfy
\be\label{two_complex_str}
\nabla_\pm I_\pm = 0 \quad,\quad \nabla_\pm = \nabla \pm g^{-1}H\;,
\ee
where to avoid overloading the notation, we have suppressed the tensor indices. As shown in \cite{Gualtieri1}, this set of data is equivalent to a pair of commuting generalized complex structures $({\cal I}, {\cal J})$, which satisfy the further property that $-{\cal I} \cdot {\cal J}$ is positive definite as a metric on $T \oplus T^{*}$ and we shall refer to the manifold equipped with this data as generalized K\"ahler.

This type of geometry enters the discussion of topological string theory in the following way: In \cite{kapustin,Kapustin:2004gv}, it was argued that a deformation of the B-model by an element of $ H^0(\Lambda^2 T^{1,0})$ corresponds to having a block-upper-triangular structure generalized complex structure ${\cal I}$. In particular, writing the corresponding pair of complex structures as $I_\pm = I \pm \delta$,  we have \cite{Gualtieri1}
$$
{\cal I} = \bem I'& \delta\cdot g^{-1}  \\ 0&  - I'^t
 \eem\quad,\quad I'=I+ {\cal O}(\delta) \;.
$$
As we have seen in the previous subsection, the upper-right corner is given by a holomorphic Poisson bi-vector of type $(2,0)$ with respect to the complex structure $I'$.

The fact that this corresponds to a non-commutative deformation can be seen from the open string sector \cite{cattaneof,deformation}, with the corresponding star product given by the Poisson vector
$$\th = \delta\cdot g^{-1} \cdot I' = \delta\cdot g^{-1} \cdot I + {\cal O}(\delta^2) \;.
$$

Using
$$
\nabla \delta= -\,g^{-1} \cdot H \cdot I \quad,\quad \nabla I= - \,g^{-1} \cdot H \cdot\delta\;,
$$
it is now straightforward to see that
$$
\nabla \th \sim g^{-1} \cdot H \cdot  g^{-1} + {\cal O}(\delta^2)\;,
$$
which is of course nothing but the same relation (\ref{poisson_H_flux}) upon proper anti-symmetrization of the indices. This provides further evidence that in the weak field limit, the relation we propose  (\ref{poisson_H_flux}) should be true for general three-form fluxes $H$.

\section{Yukawas in Non-Commutative Geometry}\label{sec:NCYuk}

In the previous section we argued that the presence of $H$-flux in F-theory induces a non-commutative deformation in the open topological string sector. We now study the effect of this non-commutative deformation on the eight-dimensional theory defined in a neighborhood containing the Yukawa enhancement point of a configuration of seven-branes.

As in section \ref{Yukawas in Commutative Geometry}, we first study the profile of the matter field zero modes, and then compute the triple overlap integral of these wave-functions. In the presence of the non-commutative deformation, the matter fields are no longer strictly localized on holomorphic curves. Recall that in the commutative case, the matter curves are identified with the loci of gauge enhancement. When the non-commutative deformation is turned on, this relation only holds to \textit{leading order} in an expansion in the deformation parameter of the theory. Indeed, although the zero modes still exhibit a Gaussian profile, as we will see in section \ref{The Localized Zero Modes}, the wave-functions now acquire more complicated extra dependence on both the holomorphic and anti-holomorphic coordinates of the local patch. Similar wave-function distortion can also be induced by gauge field fluxes, and it is therefore important to determine whether this distortion also filters through to the Yukawas.

We find that the non-commutative deformation generates hierarchical corrections to the structure of the Yukawa couplings. To establish this, we first derive a residue-like formula which makes the quasi-topological nature of the Yukawa coupling manifest. This general formalism also allows us to deduce the selection rule determined by the local rephasing symmetry of the geometry, which makes the structure of such hierarchical corrections manifest. As a check on this result, we next present an explicit example in which the Yukawa matrix as a perturbative expansion is computed.

The rest of this section is organized much as in section \ref{Yukawas in Commutative Geometry}. First we discuss localized zero modes and then establish a cohomology theory for them. After this, we obtain a residue-like formula for these Yukawa couplings, and then provide an explicit example which confirms the presence of hierarchical Yukawa structures.

\subsection{Deforming the Field Theory}

As shown in section 4, the presence of background $H$-flux in F-theory
descends to a non-commutative deformation of the seven-brane superpotential so that
the local coordinates $x$ and $y$ satisfy the relation:
\begin{equation}
x\star y - y\star x = \hbar_{\theta} \theta.
\end{equation}
Here, $\hbar_{\theta}$ is to be viewed as a quantum expansion parameter, and the two index anti-symmetric tensor $\theta=\th^{xy}$ corresponds to a local presentation of the holomorphic Poisson bi-vector:
$$
\hat \theta = \theta(x,y)\,\pa_x\wedge \pa_y\;.
$$
For constant $\theta$ and holomorphic functions $f$ and $g$, the product $f \star g$ corresponds to the usual Moyal product:
\begin{equation}
f \star g = \exp(\hbar_\th\,\theta^{jk}\partial_{\zeta^j}\partial_{\xi^k})f(z + \zeta)g(z + \xi)|_{\zeta = \xi = 0} = fg + \hbar_\th\,\theta^{jk} \partial_{j} f \partial_{k} g + O(\hbar_\th^2).
\end{equation}
For general, non-constant $\th$, the $\star$-product is given in (\ref{starproduct}). To keep our discussion as general as possible, we shall consider $\theta$ non-constant. In particular, we shall later see in section 6 that the case where $\theta$ vanishes at the classical Yukawa point leads to phenomenologically interesting Yukawa structures.

To write down the non-commutative version of the superpotential, it is necessary to extend the action of $\star$ to differential forms as well. We shall refer to the non-commutative deformation of the wedge product as $\ost$. Utilizing the properties of the $\ost$ product detailed in Appendix B, it is possible to show that the $\ost$-product is the only available deformation of the $\wedge$ product which satisfies the following properties:
\begin{enumerate}
\item{Associativity}
\item{The $\bar\pa$-chain rule}
$$
\bar\pa (F \ost F') = \bar\pa F \ost F' +(-1)^{\text{deg}F} F  \ost\bar\pa F'\quad \quad\text{for $(p,q)$-forms }F,F'
$$
\item{Commutativity Upon Integration}
\be\label{SWcondition}
\int_S \Tr \big( F\ost F' \big) =\int_S \Tr \big( F'\ost F \big) \;,
\ee
\end{enumerate}
for forms $F$ and $F'$ which satisfy an appropriate notion of localization. As discussed in \cite{seibergwitten}, these properties are all crucial in order for the corresponding theory to be gauge invariant. We refer to Appendix \ref{Proof of Commutativity Upon Integration} for the proof of property (\ref{SWcondition}).

The non-commutative deformation of the seven-brane superpotential for gauge group $G$ is then given by:
\be\label{superpotential_NC}
W = \int_S \Tr \Big( \,\big( \bar\pa \bar A + \bar A \ost \bar A \big) \ost \varphi \Big)\;.
\ee
Here, it is important to note that in the presence of the non-commutative deformation and for a general gauge group $G$, the fields $\bar A$ and $\varphi$ will not in general take values in the adjoint representation of $G$, but will instead take values in the universal enveloping algebra of $G$ \cite{bonora, wess}. Due to the associative structure of the group $U(N)$, it follows that in this special case, this subtlety is absent. To simplify our analysis, we will therefore confine our discussion to this choice of gauge group. Nevertheless, much of the general methodology discussed throughout this paper is likely also applicable to general gauge groups, and in particular to $E$-type gauge groups which have proven to be quite important in the context of F-theory GUTs \cite{Heckman:2009mn}.

With the above definition of the non-commutative $\ost$-product, the superpotential (\ref{superpotential_NC}) is indeed a deformation of the original theory (\ref{superpotential}) which reduces to the original one when the deformation parameter $\theta$ vanishes.
As promised, this deformed superpotential is invariant under the gauge
transformation:
\be\label{gauge_inv_NC}
\bar A \to g^{-1_\ost} \ost\bar A  \ost  g +g^{-1_\ost}   \ost \bar\pa g\quad,\quad \varphi \to
g^{-1_\ost} \ost \varphi   \ost g\;,
 \ee
where $g^{-1_\ost}$ is the inverse of $g$ under $\ost$:
$$
g^{-1_\ost}\ost {\mathit g} = g\ost g^{-1_\ost} = 1\;.
$$
Here, the formal inverse $g^{-1_\ost}$ is defined in terms of
an expansion in $\hbar_{\theta}$. So long as $g^{-1}$ exists, it can be shown that
a unique $g^{-1_\ost}$ can be constructed recursively.

For later use we also write down the infinitesimal version of the above gauge transformations:
\be\label{infini_gauge_trnsf_NC}
\d \bar A = \bar\pa f + [\bar A, f]_\ast \quad,\quad \d \varphi = [\varphi, f]_{\ast} \;,
\ee
where we have introduced the notation
$$
[F,F']_\ast = F\ost F' -
F' \ost F
$$
for later convenience.

Varying the superpotential of equation \eqref{superpotential_NC} with respect to $\bar A$ and $\varphi$ yields the deformed F-term equations of motion:
\bea\notag
F^{(0,2)}_\ost= \bar\pa \bar A + \bar A \ost \bar A &=&0 \\ \label{F_term_eq_NC}
\bar\pa \varphi + [\bar A,\varphi]_\ast &=& 0
\eea
which are given by replacing all wedge products by $\ost$-products in the undeformed equations of motion.
From the gauge invariance of (\ref{gauge_inv_NC}), we see that it is again possible to choose the holomorphic gauge $\bar A =0$ in the deformed theory for a supersymmetric field configuration without any loss of generality.

\subsection{Matter From Non-Commutative Geometry}

In this subsection we repeat our analysis of localized zero modes done in the case of the commutative theory, but now in the presence of the non-commutative deformation. To this end, we first study solutions to the background field F-term equations of motion, which reduce in the limit $\hbar_{\th} \rightarrow 0$ to those of the commutative theory. Next, we study the profile of zero modes which are localized on ``fuzzy'' matter curves. Just as in the
undeformed case, the gauge equivalence of the localized zero modes is captured by an appropriately defined cohomology, a
fact which will play an important role in our analysis of the Yukawa coupling between these localized zero modes.

\subsubsection{Background Field Configurations}
\label{The Matter Curves NC}

First we will specify the supersymmetric background configuration $\anull, \fnull$ of the non-commutative topological field theory in which we would like to study the ``quantum Yukawa coupling'' of the non-commutative theory.

As mentioned earlier, to keep our analysis of the non-commutative theory as simple as possible, we shall confine our attention to the case of a $U(n)$ non-commutative gauge theory so that the background fields, and the corresponding zero modes are given by $\mathfrak{u}(n)$-valued differential forms. It will often be useful to think of them as $n\times n$ matrices and denote different modes as distinct entries in this matrix.

Much as in the commutative theory, in the non-commutative theory we can choose a holomorphic gauge with
\be
\anull = 0 \;.
\ee
In this case the other F-term equation (\ref{F_term_eq_NC}) again implies that $\fnull$ is a matrix with holomorphic entries.
As before, the most natural choice for such a background is one which satisfies:
$$
[\fnull, \bar\varphi^{(0)}]_{\ast}=[\fnull, \bar\varphi^{(0)}]=0 \;.
$$
For this reason, we shall again focus on backgrounds where $\fnull$ takes values in the Cartan of $\mathfrak{u}(n)$ and can be represented in matrix form as:
\be
\fnull = \finull dx\wedge dy = \text{diag}(\finull_1,\dotsi,\finull_n) \,dx\wedge dy \;,
\ee
where the $\finull_i(x,y)$'s are some holomorphic sections of the bundle $K_S|_U$.

An important feature of the above definition of $\finull$ is that the overall
trace $\finull_1 + ... + \finull_n$ is not subject to any constraint, since
the matrix $\finull$ takes values in $\mathfrak{u}(n)$. Indeed, as we will show later in this section,
this overall trace generates an additional expansion for hierarchical Yukawa structures.

It might at first appear that the absence of the trace condition is in tension with the
expectation that the Yukawa enhancement point defines a gauge theory of $ADE$ type.
Indeed, in the commutative case, the associated matrices $\finull$ are typically traceless with
respect to the Lie algebra for the gauge group defined at the Yukawa enhancement point. We have
already mentioned that in the non-commutative gauge theory, the fields take values
in the universal enveloping algebra, and so will typically not be traceless.

Aside from this formal expectation, there is another more direct reason to expect the corresponding
trace to be important in many cases of interest. Returning to the undeformed field theory, note that
although $\finull$ will satisfy an overall trace condition, any $m \times m$ sub-block
for $1 \leq m \leq n$ of $\finull$ need not satisfy such a constraint. In particular, the overall trace
of a sub-block can then enter into the Yukawa overlap integral. This is especially natural in the context of higher rank Yukawa enhancement points, as has been advocated for example in \cite{Heckman:2008qa,BHSV,Heckman:2009mn}.

\subsubsection{Fuzzy Matter Curves and Localized Zero Modes}
\label{The Localized Zero Modes}

Having specified a class of background field configurations which closely mimic those of the commutative gauge theory,
we now study localized zero mode solutions in this background. In the limit where $\hbar_{\theta} \rightarrow 0$, we recover the expected loci of gauge enhancement corresponding to the matter curves $\Sigma_{ab} $:
$$
\Sigma_{ab}= \Sigma_{ba} = \{\finull_a- \finull_b=0\}\quad,\quad a, b = 1,\dotsi , n\;.
$$
For this reason we will sometimes refer to the $\Sigma_{ab}$ as the ``classical'' matter curves. Including the
effects of the non-commutative deformation, the precise notion of where the zero modes are localized becomes ``fuzzy''.

The equations of motion the zero modes satisfy are obtained by separating the fields into the background ($\anull, \fnull$) and the deformation ($\aone, \fone$)  part as before and expanding the deformed F- and D-term equations of motion to linearized order. The resulting equations are again simply given by replacing all wedge products in the original zero mode equations (\ref{zero_mode_fterm_eq1}-\ref{zero_mode_fterm_eq2})
by $\ost$-products and read
\bea\notag
\bar \pa \aone + \anull \ost \aone +\aone\ost\anull &=& 0\\ \label{NC_z_mode_eq}
\bar \pa \fone + [\anull,\fone]_\ast + [\aone,\fnull]_\ast &=& 0 \;.
\eea
for the F-term equations of motion. The D-term equations given by
the analogue of equation (\ref{dterm_zeromodes}) read as:\footnote{Although we have argued
in section 4 that $H$-fluxes induce a non-commutative deformation of the F-term equations
of motion, the D-term equations of motion are strictly speaking outside the realm of the topological B-model, and
so cannot be captured in a similar fashion. Note, however, that D-terms play a secondary role in the present case, because we shall mainly
be interested in deformations of the superpotential. It is therefore
enough to specify the K\"ahler data such that it is in principle compatible with the
gauge invariance of the superpotential. For example, our definition of
the $\ost$ product acts trivially on anti-holomorphic coordinates.}
\be\label{Dterm_NC}
\omega \ost \big(\bar\pa \aone + A^{(0)}  \ost \aone + \aone \ost  A^{(0)} \big)
+ \frac{i}{2} [\bar \varphi^{(0)} , \fone]_\ast = 0 \;.
\ee
As before we shall refer to zero modes which satisfy both the F- and D-term equations of motion as the ``physical'' zero mode wave-functions.

We now solve for the zero modes in the presence of the background field configuration defined in section \ref{The Matter Curves NC} where $\fnull$
takes values in the Cartan subalgebra of $\mathfrak{u}(n)$. The analysis closely parallels the undeformed case
presented in equation (\ref{z_modes_2}). The solutions to the zero mode equation (\ref{NC_z_mode_eq}) can be written in the following form
\be\label{NC_z_mode_1}
\aone = \bar\pa \xi \quad,\quad \fone = [\fnull,\xi ]_\ast + h
\ee
with some matrix-valued regular function $\xi$ and holomorphic section $h$ of the canonical bundle. In matrix components they read
\be\label{NC_z_mode_2}
\aone_{ab} = \bar \pa \xi_{ab}\quad,\quad \fione_{ab} = \finull_a \ost \xi_{ab} - \xi_{ab}\ost \finull_b + h_{ab} \quad \text{no sum over  }a,b\;,
\ee
where, as we recall from the definition of (\ref{ostar})
$$
\finull_a \ost \xi_{ab}  =\big( (\theta\finull_a)\star \xi_{ab} \,\big) \,\theta^{-1}\;.
$$
As opposed to the commutative case, note that the holomorphic section $h_{ab}$ and the
zero mode wave-function $\phi^{(1)}_{ab}$ will in general only agree on the classical
matter curve $\Sigma_{ab}$ in the strict limit $\hbar_{\theta} \rightarrow 0$.

We are interested in zero mode solutions which vanish rapidly away from the classical matter curves $\Sigma_{ab}$. With the gauge equivalence (\ref{infini_gauge_trnsf_NC}) taken into account, we conclude that, on our patch, the space of gauge inequivalent classes of localized
zero mode solutions is isomorphic to the space of holomorphic sections, with a now deformed equivalence relation
$$
h_{ab}\sim h_{ab} + \finull_a \ost \til h_{ab} - \til h_{ab}\ost \finull_b\;,
$$
for any functions $\til h_{ab}$.

Stated in another way, for a given holomorphic section $h_{ab}$  one can find holomorphic functions on the patch $U\setminus \Sigma_{ab}$ such that
\be\label{def_zeta}
h_{ab}= \finull_a \ost \zeta_{ab} - \zeta_{ab}\ost \finull_b\quad {\text{on   }}U\setminus \Sigma_{ab}\;.
\ee
The smooth function $\xi_{ab}$ reduces to $-\zeta_{ab}$ far away from the matter curve $\Sigma_{ab}$ due to the localization of the wave-function $\fione_{ab}$. The two zero modes given by the holomorphic functions $h_{ab}$ and $h'_{ab}$ are gauge equivalent if and only if the corresponding $\zeta_{ab}$ and $\zeta'_{ab}$ differ by a function that can be holomorphically extended to the entire patch $U$.

Furthermore, we can define a deformed version of the $\bar D$ operator introduced in section \ref{The Cohomology Theory of Chiral Matter}
\be
\bar D_\ast = \bar \pa + \hat \k \wedge [\finull, \cdot\;]_\ast\;.
\ee
In the matrix notation, where we define $E_{ab}$, $a,b = 1,\dotsi,n$  to be the $n\times n$ matrix with $1$ in the $a$-$b$-th entry and zero for all other entries, it reads
\be
\bar D_\ast \big( (F_{ab} + \hat \k \wedge G_{ab}) \,E_{ab} \big) = \Big( \bar\pa F_{ab} - \hat\k\wedge
\big( \bar\pa G_{ab} +F_{ab} \ost \finull_b - \finull_a \ost F_{ab}\big)\, \Big) \,E_{ab}\;,
\ee
where $F_{ab}$ is a $(0,k)$-form and $G_{ab}$ a $K_S$-valued $(0,k-1)$-form on the surface $S$ as before.

It is not difficult to check that the deformed $\bar D_\ast$ operator still has the
following properties:
\bea\notag
\bar D_\ast^2 &=& 0 \\
 \bar D_\ast (\Psi \ost \Psi') &=& \bar D_\ast \Psi \ost \Psi' + (-1)^{\text{deg}\Psi}  \Psi \ost \bar D_\ast \Psi' \; \;,
\eea
and that the zero mode equation and the gauge invariance of the theory (\ref{superpotential_NC}) can now be written in the cohomological form
\be
\bar D_\ast \Psi = 0 \quad,\quad \Psi \sim \Psi + \bar D_\ast \,f
\ee
for any matrix-valued, smooth function $f$ subject to some support properties.
Finally, for future use we note that the general form of the zero mode solutions (\ref{NC_z_mode_1}) can be conveniently rewritten as
\be\label{z_mode_Psi_NC}
\Psi^{(1)} = \aone + \hat\k\wedge \fione = \bar D_\ast \,\xi + \hat\k \wedge h \quad,\quad \bar\pa h = 0 \;.
\ee

\vspace{18pt}
{\it{An Example}}
\vspace{12pt}

To illustrate the effects of the non-commutative deformation on the profile of the zero mode wave-functions,
we now present an example in which we solve the F- and D-term equations of motion.
For simplicity, we choose a canonical K\"ahler form, so that in local coordinates:
$$
\o = \frac{i}{2} g \big(dx \wedge d\bar x +dy \wedge d\bar y \big),
$$
and assume that the Poisson bi-vector $\hat \theta$ is constant on the patch. We consider a configuration in which the
background gauge field strength is switched off
$$
A^{(0)} = 0\;,
$$
and in which $\finull$ is diagonal. In this case the
F-term (\ref{F_term_eq_NC}) and D-term (\ref{Dterm_NC}) equations, in matrix notation, become
\bea\notag
\bar\pa \aone_{ab} &=& 0\\\notag
\bar\pa \fione_{ab} + \aone_{ab}\ost \finull_b - \finull_a \ost  \aone_{ab}&=& 0 \\
g \, \big( \pa_x \aone_{ab,\bar x}+ \pa_y \aone_{ab,\bar y}\big) - (\bar \f^{(0)}_a-\bar \f^{(0)}_b)\,\fione_{ab}&=&0\;.
\eea

Now take $\finull$ in the Cartan of $\mathfrak{u}(2)$ so that:
$$
\fnull = -\frac{1}{2}\bem x+C & 0 \\ 0& -x+C\eem dx\wedge dy \;,
$$
where $C$ is a complex constant. The zero mode solution in the 1-2 entry of the matrix is
\bea\notag
\aone_{12} &=& \frac{1}{\sqrt{g}} \, e^{-\frac{1}{2 \sqrt{g}} \bar x(x+C)}\ost h_x(y) \ost e^{-\frac{1}{2 \sqrt{g}} \bar x(x-C)} d\bar x \\
\fione_{12} &=& e^{-\frac{1}{2 \sqrt{g}} \bar x(x+C)}\ost h_x(y) \ost e^{-\frac{1}{2 \sqrt{g}} \bar x(x-C)}\;,
\eea
where $h_x(y)$ is a holomorphic function of the coordinate $y$.

In this setup, the wave-functions are simply the ``normal-ordered'' versions of the wave-functions (\ref{commutative_simple_wavefunction}) of the undeformed theory. Furthermore, as long as the non-commutativity parameter $\th$ does not vanish, the wave-functions cannot be deformed to a $\d$-function anymore, no matter how we tune the K\"ahler data $g$. This is to be
contrasted with the commutative case (\ref{localized_delta_zero_mode}) and suggests that the triple coupling does not take place just strictly at a point anymore but rather in a ``fuzzy neighborhood'' of the classical Yukawa point. We now show that this allows the non-commutative deformation to violate the rank theorem of section \ref{A Rank Theorem}.

\subsection{The Yukawa Coupling}

In this sectioned we study the effects of the non-commutative deformation on the structure of the Yukawa. We find that on the one hand, a very similar residue-like formula can be obtained and in particular, the deformed Yukawa coupling is still a topological quantity in the sense that it does not depend on the representative in the $\bar D_\ast$-cohomology one picks for the zero mode wave-functions. On the other hand, we also find that the rank theorem discussed in section \ref{A Rank Theorem} for the undeformed theory does not hold when the
non-commutative deformation is activated. In particular, we find that the non-commutative deformation generates hierarchical corrections to the Yukawas.

\subsubsection{A Quantum Residue Formula and a Selection Rule}

In the deformed superpotential (\ref{superpotential_NC}), substituting $\bar A = \anull+\aone$ and $\varphi = \fnull + \fone$ we obtain the Yukawa coupling
\be
W_{\text{Yuk}} =  \int_U \Tr \big(\aone\ost \aone\ost \fone \big)
\ee
as the only potentially non-vanishing piece of the superpotential. In the language of the twisted one-forms (\ref{z_mode_Psi_NC}) we again obtain the equivalent formula
\be
W_{\text{Yuk}} =  \int_U \Tr \big(\Psi^{(1)}\ost \Psi^{(1)}\ost \Psi^{(1)} \big)
\ee
where the rule of integration is given by (\ref{twisted_integral}) as before.

To make the discussion more explicit and without loss of generality, we focus on the contribution:
$$
W_{\text{Yuk},123} = \int_U  \big( \Psi^{(1)}_{12}\ost \Psi^{(1)}_{23}\ost \Psi^{(1)}_{31}  \big)
$$
to the above Yukawa coupling. It is clear that such a term can be discussed independently from other possible terms. We will hence drop the subscript `123' in $W_{\text{Yuk},123}$.

To evaluate this overlap integral, we follow the same general strategy outlined in section \ref{Yukawa Coupling as Residue} and Appendix A. To be explicit, we shall take the patch $U$ to be the bi-disk:
$$
U = D_x\times D_y = \{x: |x|\leq R\} \times \{y: |y|\leq R\}\;,
$$
with the coordinates chosen such that the classical matter curves are given by $\finull_{1} -\finull_2=-x=0\,$ and  $\,\finull_{2}-\finull_3=y=0$. From here we immediately see that automatically there is a third classical matter curve $\finull_{3} -\finull_1=x-y=0$ passing through the same intersection point. In other  words, to study the Yukawa coupling associated to a specific intersection point, we will work with a $U(3)$ gauge group with the background
\be\label{U3_bkgnd_NC}
\fnull = \frac{1}{3} \bem -2x+y+\Phi &0&0 \\ 0&x+y+\Phi & 0 \\ 0&0& x-2y+\Phi \eem\,dx\wedge dy\;,
\ee
where $\Phi = \Phi(x,y)$ is a holomorphic function.

Working on the patch $U$ and following the analysis in section \ref{Yukawa Coupling as Residue} and Appendix A, we obtain:
\bea\notag
W_{\text{Yuk}} &=& \int_U \big( \Psi^{(1)}_{12}\ost \Psi^{(1)}_{23}\ost \bar D_\ast\, \xi_{31}  \big)+
 \int_U  \Psi^{(1)}_{12}\ost \Psi^{(1)}_{23}\ost \big( h_{31}  \,dx\wedge dy\big)\\\notag
 &=&  \int_U \bar\pa \xi_{12}\ost \bar\pa \xi_{23}\ost\big(  h_{31}  \,dx\wedge dy\big)\\
 &=& \oint_{|x|=R}\,\oint_{|y|=R}  \xi_{12}\ost \xi_{23}\ost  \big( h_{31}  \,dx\wedge dy\big)\;,
\eea
where we have dropped the first term in the first line because it can be written as a boundary term:
$$
\int_U \big( \Psi^{(1)}_{12}\ost \Psi^{(1)}_{23}\ost \bar D_\ast\, \xi_{31}  \big)= \int_{\pa U}\big( \Psi^{(1)}_{12}\ost \Psi^{(1)}_{23}\ost  \xi_{31}  \big)
$$
which corresponds to an exponentially suppressed contribution because $\Psi^{(1)}_{12}\ost \Psi^{(1)}_{23}$ vanishes rapidly away from the origin, assuming again that the matter curves $\Sigma_{12}$ and $\Sigma_{23}$ intersect transversely.

Finally, from the fact that the functions $\xi_{ab}$ reduce to the holomorphic functions $-\zeta_{ab}$ away from the corresponding matter curves due to the localization of the wave-functions (\ref{def_zeta}), for large $R$ we can further rewrite the above expression as
\be
W_{\text{Yuk}} =  \oint_{|x|=R}\,\oint_{|y|=R}  \zeta_{12}\ost \zeta_{23}\ost\big( h_{31} \;dx\wedge dy\big)\;,
\ee
with $\zeta_{ab}$ given by the holomorphic section $h_{ab}$ by the relation $h_{ab}= \finull_a \ost \zeta_{ab} - \zeta_{ab}\ost \finull_b$ which holds away from the matter curve.

Again fixing the zero mode on the third matter curve to be given by $h_{31}=1$, we get the following expression for the Yukawa coupling as the ``quantum bilinear pairing'' of the two curves
\be\label{quantum_residue}
W_{\text{Yuk}} =  \oint_{|x|=R}\,\oint_{|y|=R}   \frac{h_{12}}{[\finull,\cdot]_{12,\ast}} \ost  \frac{h_{23}}{[\finull,\cdot]_{23,\ast}} \ost \;\big(dx\wedge dy\big)\;,
\ee
where we have introduced the notation
$$
\zeta_{ab}= \frac{h_{ab}}{[\finull,\cdot]_{ab,\ast}}\Leftrightarrow h_{ab}= \finull_a \ost \zeta_{ab} - \zeta_{ab}\ost \finull_b$$
for convenience. Note that on both sides of the equation, $h_{ab}$ and $\finull_a$, $\finull_b$ are all sections of the canonical bundle $K_S$, and hence it is easy to see that the quantity $\zeta_{ab}$ is indeed a holomorphic function away from the matter curve.
This equation can be thought of as the quantized version of the residue formula (\ref{eq:matrixY}) for the undeformed theory and indeed reduces to the residue formula when the limit $\th\to0$ is taken.

To see whether the rank theorem also applies to the deformed theory, we now take a closer look at the object
$\zeta_{12}$
which enters the quantum residue formula (\ref{quantum_residue}).
Define the bi-differential operator $\ost_k, k=0,1,\dotsi$ as the coefficients of the $\hbar_{\theta}$-expansion of the operator $\ost$, such that
$$
f \ost g = \sum_{k=0}^\inf \hbar_{\theta}^k \, ( f \ost_k g\,)\;,
$$
then $\zeta_{12} = \sum_{k=0}^\inf \hbar_{\theta}^k \zeta_{12,k}$ is given by the following recursive relation
\be\notag
\z_{12,0} = \frac{h_{12}}{\finull_1-\finull_2} = - \frac{h_{12}}{x} \quad,\quad \sum_{k=0}^n \finull_1\ost_k \z_{12,n-k}- \z_{12,n-k} \ost_k \finull_2=0 \quad\text{for all }n>0\;.
\ee

At next order, the second equation gives
\bea\notag
\zeta_{12,1} &=& -\frac{1}{x} \Big(  (\finull_1 + \finull_2 )\ost_1 (\frac{h_{12}}{x})\Big) \\\label{invert_ost}
&=&  -\frac{1}{x} \Big( V_1(\theta\finull_1 + \theta\finull_2 )\, V_2(\frac{h_{12}}{x})-V_2(\theta\finull_1 + \theta\finull_2 )\, V_1(\frac{h_{12}}{x})\Big)\,\theta^{-1}
\eea
where $V_{1,2}$ are the differential operator defined by the Poisson bi-vector (\ref{def_V}). And the same structure also holds for $\zeta_{23}$ of course.

From the above expression and the quantum residue formula (\ref{quantum_residue}) we can immediately see a $U(1)$ selection rule which is very reminiscent of the Froggatt-Nielsen mechanism. A necessary condition for a term in the integrand in the Yukawa quantum residue formula (\ref{quantum_residue}) to contribute is that it has total charge zero under the geometric $U(1)$
$$
x\to e^{i\th} x\quad,\quad y\to e^{i\th} y\;.
$$
Expanding $\theta(x,y)$ as:
\begin{equation}
\theta(x,y) = \sum_{m_1+m_2\geq \ell} \theta_{m_1,m_2} x^{m_1} y^{m_2}
\end{equation}
for some $\ell \geq 0$, it follows that each $\theta_{m_1,m_2}$ can be viewed as having a charge under this $U(1)$ rephasing symmetry. In particular, $\theta_{m_1,m_2}$ has charge $m_1 + m_2 - 2$. The order of vanishing of $\theta$ can therefore have a significant impact on the possible Yukawa textures expected.

Similar considerations apply to $\Phi(x,y) = \Tr\big(\finull\big)$:
\be
\Tr\big(\finull\big) = \Phi(x,y) = \Phi_0 + \sum_{m_1+m_2\geq 0} \Phi_{m_1,m_2}\,  x^{m_1} y^{m_2}\;.
\ee
In this case, we find that the terms $\Phi_{m_1,m_2}$ have charge $m_1 + m_2 -1$ relative to $\finull_a - \finull_b$ under the geometric $U(1)$. In particular, the only term of $\Phi$ which carries a negative charge relative to $\finull_a - \finull_b$ is $\Phi_{0}$, with charge $-1$.

From this analysis, we obtain the conclusion that, when using the monomials $y^m$, $x^n$ as a basis for the holomorphic functions $h_{12}(y)$, $h_{23}(x)$ labelling the zero modes, a necessary condition for the Yukawa coupling
\bea\notag
W_{\text{Yuk}}^{mn} =  \oint_{|x|=R}\,\oint_{|y|=R} \frac{y^m}{[\finull,\cdot]_{12,\ast}}\ost   \frac{x^n}{[\finull,\cdot]_{23,\ast}} \;\ost \big(dx\wedge dy\big) =\sum_{\substack{k_1,k_2\in\Z_+ \;,\;k_1\geq k_2\\  k_1(2-\ell) +k_2 \geq m+n} } T_{k_1,k_2}\,\hbar_{\theta}^{k_1} \Phi_0^{k_2}
\eea
to contain a term of the form $\hbar_{\theta}^{k_1} \Phi_0^{k_2}$ is
$$
k_1\geq k_2\quad,\quad  k_1(2-\ell) +k_2 \geq m+n\;,
$$
where $\ell$ is the minimal degree of the polynomial $\th$ in holomorphic coordinates $x$, $y$.

\vspace{18pt}
{\it{An Example}}
\vspace{12pt}

Above we have derived a geometric $U(1)$ selection rule which naturally implements the Froggatt-Nielsen mechanism that renders a hierarchical structure in the Yukawa matrix.  To illustrate it, let us take $\ell=1$, namely let us choose our Poisson bi-vector to have the expansion
$$
\theta(x,y) = \theta_x \,x + \theta_y \, y + \text{higher power}
$$
and similarly
$$
\Tr\big(\finull\big) = \Phi(x,y) dx \wedge dy = (\Phi_0+ \text{higher power}) dx \wedge dy \;.
$$
For example, in a three-generation model, the Yukawa coupling matrix can now be expressed in terms of a series expansion in $\hbar_{\theta}$ and $\Phi_0$ of the form
\be\label{selection_NC_example}
\big(W_{\text{Yuk}}\big) \sim \bem 1&\hbar_{\theta} &\hbar_{\theta}^2 \\ \hbar_{\theta} &\hbar_{\theta}^2 &\hbar_{\theta}^3 \\\hbar_{\theta}^2 &\hbar_{\theta}^3 &\hbar_{\theta}^4 \eem +  \Phi_0\,\bem 0&0 &\hbar_{\theta} \\ 0 &\hbar_{\theta} &\hbar_{\theta}^2 \\\hbar_{\theta} &\hbar_{\theta}^2 &\hbar_{\theta}^3 \eem +
\Phi_0^2 \bem 0&0&0\\0&0&0 \\ 0&0& \hbar_{\theta}^2\eem
\ee
for the scaling of the leading terms in the expansion in $\hbar_{\theta},\Phi_0 \ll 1$.

\subsubsection{An Explicit Example}

To illustrate the structure of the deformed Yukawa coupling, we now present a simple explicit example.
In the background
\be
A^{(0)} = 0 \;\;,\;\;\fnull=  \frac{1}{3} \bem -2x+y+\Phi_0 &0&0 \\ 0&x+y+\Phi_0 & 0 \\ 0&0& x-2y+\Phi_0 \eem\,dx\wedge dy\;, \;\Phi_0 \in \C\;,
\ee
take the Poisson bi-vector to be
$$
\hat \theta = 2\,(x-y) \,\pa_x\wedge \pa_y = \, z \pa_z \wedge \pa_w\;,
$$
where
$$
z= x-y\quad,\quad w = x+y\;,
$$
then the solution to
$$
\finull_1 \ost \zeta_{12} - \zeta_{12}\ost \finull_2 = h_{12} \;\;,\;\;\finull_2 \ost \zeta_{23} - \zeta_{23}\ost \finull_3 = h_{23}
$$
is given by
\bea\notag
\zeta_{12}&=&\sum_{n=0}^\inf \hbar_{\theta}^n \zeta_{12,n}\;,\; \zeta_{12,0} = \frac{-2}{z+w}\,h_{12}\;,\;\zeta_{12,n} = \frac{-2}{z+w}\sum_{m=1}^n \frac{1}{m!} D_{m}^{(12)}\,\zeta_{12,n-m}\\\notag
\zeta_{23}&=&\sum_{n=0}^\inf \hbar_{\theta}^n \zeta_{23,n}\;,\; \zeta_{23,0} = \frac{-2}{z-w}\,h_{23}\;,\;\zeta_{23,n} = \frac{-2}{z-w}\sum_{m=1}^n \frac{1}{m!} D_{m}^{(23)}\, \zeta_{23,n-m}\\\notag
D_{m}^{(12)}&=&\big(\frac{1}{6}+\frac{(-1)^m}{3}\big)\,\big( w\pa_w^m -m\,z\,\pa_z\pa_w\big) +2^{m-1} \,z\,\pa_w^m + \frac{\Phi_0}{3} \big(-1 + (-1)^m\big)\pa_w^m\\\notag
D_{m}^{(23)}&=&-\big(\frac{(-1)^m}{6}+\frac{1}{3}\big)\,\big( w\pa_w^m -m\,z\,\pa_z\pa_w\big) -(-2)^{m-1} \,z\,\pa_w^m + \frac{\Phi_0}{3} \big(-1 + (-1)^m\big)\pa_w^m
\eea

Taking the zero modes to be given by
$$
h_{12,m} = \frac{y^m}{m!} \quad,\quad h_{23,n} = \frac{x^n}{n!} \;,
$$
for $m,n = 0,1,2$, the leading order behavior of the Yukawa matrix we obtain by evaluating (\ref{quantum_residue}) is:
\be\notag
\frac{1}{(2 \pi )^{2}}W_{\text{Yuk}} =
\bem 1&\frac{4}{3}\hbar_{\theta} &\frac{11}{9}\hbar_{\theta}^2 \\ -\frac{4}{3}\hbar_{\theta} &-\frac{26}{9}\hbar_{\theta}^2 & -\frac{107}{27}\hbar_{\theta}^3 \\ \frac{11}{9}\hbar_{\theta}^2 &\frac{107}{27} \hbar_{\theta}^3 &\frac{1225}{162}\hbar_{\theta}^4 \eem +  \Phi_0\,\bem 0&0 &-\frac{1}{6}\hbar_{\theta} \\ 0 &0 &
\frac{2}{9}\hbar_{\theta}^2 \\ \frac{1}{6}\hbar_{\theta} &\frac{2}{9}\hbar_{\theta}^2 &0 \eem +
\Phi_0^2 \bem 0&0&0\\0&0&0 \\ 0&0& -\frac{1}{36}\hbar_{\theta}^2\eem\;,
\ee
which is indeed of the form (\ref{selection_NC_example}) with order one coefficients.

\section{Applications \label{sec:App}}

In this section we make contact between the formal Yukawa structures found in
previous sections, and more phenomenological applications. To this end, we
first review the proposal of \cite{Heckman:2008qa} that background fluxes can
distort the structure of the Yukawa couplings. We next show how
quite similar expressions to those estimated in \cite{Heckman:2008qa}
appear in the non-commutative deformation of the
partially twisted seven-brane theory.

\subsection{Review of FLX and DER Expansions \label{RevFLX}}

We now briefly review the proposal of \cite{Heckman:2008qa} that background
fluxes can generate hierarchical corrections to the structure of the Yukawa couplings. Letting $z_{\Sigma_{i}}$ denote a local coordinate along a matter
curve $\Sigma_{i}$ such that $z_{\Sigma_{1}}=z_{\Sigma_{2}}=z_{\Sigma_{3}}=0$
denotes the interaction point, there exists a basis of wave-functions
organized by the order of vanishing near the interaction point so that:%
\begin{equation}
\Psi_{\Sigma}^{(i)}\sim\left(  z_{\Sigma}\right)  ^{i}+O(\left(  z_{\Sigma
}\right)  ^{i+1}),
\end{equation}
where the localized matter fields exhibit an exponential falloff in
directions normal to the matter curve.

The leading order profile of the wave-functions exhibits a rephasing symmetry
due to the local action of the internal Lorentz symmetries on the holomorphic
coordinates $z_{\Sigma}$. This overall rephasing symmetry implies that overlap
integrals of the form:%
\begin{equation}
\lambda^{ijk}\sim%
{\displaystyle\int}
\left(  z_{\Sigma_{1}}\right)  ^{i}\left(  z_{\Sigma_{2}}\right)  ^{j}\left(
z_{\Sigma_{3}}\right)  ^{k}\times Gauss
\end{equation}
will vanish unless $i=j=k=0$. Here, ``$Gauss$'' is shorthand for the rephasing
invariant exponential falloff of the wave-functions, with the precise form set by the behavior of the K\"{a}hler form near the intersection point.

The rephasing symmetry of the holomorphic coordinates determines a selection
rule for the Yukawa matrices derived from the geometry. As an example, in the
up type Yukawa coupling defined by the interaction term:%
\begin{equation}
W_{MSSM}\supset\lambda_{u}^{ij}Q^{i}U^{j}H_{u},
\end{equation}
the $3\times3$ matrix $\lambda_{u}^{ij}$ has rank one when the $U(1)$
selection rule is exactly obeyed. This would correspond to a situation with
one massive generation, and two exactly massless generations. In any
semi-realistic theory of flavor, this structure must be corrected to
incorporate hierarchical masses and mixing angles.

Subleading corrections to the form of the Yukawa coupling correspond to
violations of this rephasing symmetry. In \cite{Heckman:2008qa}, it was
proposed that distortions of the integrand by additional powers of the
anti-holomorphic coordinates $\bar{z}_{\Sigma}$ could induce higher order
corrections to the structure of the Yukawas. The first order correction to the
form of the integrand can occur from objects with one holomorphic and two
anti-holomorphic tensor indices of the form:%
\begin{equation}
\mathcal{O}_{\mathcal{F}}=\mathcal{F}_{i\bar{j}\bar k}z_{i}
{\bar z_{\bar j} \bar z_{\bar k}},
\end{equation}
so that locally $\mathcal{F}$ behaves as a $(1,2)$ form. Examples of such
tensor structures are gradients of the gauge field strength of the form
$\nabla_{i}F_{{\bar j\bar k}}$ and $\nabla_{\bar{i}%
}F_{j\bar{k}}$, as well as the $(1,2)$ component of the background
$H$-flux. As we shall argue later, these contributions are actually closely
linked in the context of the non-commutative deformation of the superpotential.

On general grounds, the rephasing symmetry can be violated by higher
order terms either of the form:%
\begin{equation}
(\mathcal{O}_{\mathcal{F}})^{M}\cdot\left(  z_{\Sigma_{1}}\right)  ^{i}\left(
z_{\Sigma_{2}}\right)  ^{j}\left(  z_{\Sigma_{3}}\right)  ^{k}\times Gauss
\end{equation}
or:%
\begin{equation}
(\bar{\partial}^{M}\mathcal{O}_{\mathcal{F}})\bar{z}^{M}\cdot\left(
z_{\Sigma_{1}}\right)  ^{i}\left(  z_{\Sigma_{2}}\right)  ^{j}\left(
z_{\Sigma_{3}}\right)  ^{k}\times Gauss.
\end{equation}
The first expansion corresponds to additional contributions in powers of the
flux, and was referred to as the \textquotedblleft FLX
expansion\textquotedblright\ in \cite{Heckman:2008qa}. The second expansion
corresponds to including higher order gradients of $\mathcal{O}_{\mathcal{F}}$
and was referred to as the \textquotedblleft DER expansion\textquotedblright.
Counting the overall $U(1)$ charge of each integrand, the resulting structure
is similar to that of the Froggatt-Nielsen mechanism \cite{Froggatt:1978nt}.
The resulting form of the two expansions leads to Yukawa
matrices of the form \cite{Heckman:2008qa}:
\begin{equation}
\lambda_{FLX}\sim\left(
\begin{array}
[c]{ccc}%
1 & \varepsilon^{2} & \varepsilon^{4}\\
\varepsilon^{2} & \varepsilon^{4} & \varepsilon^{6}\\
\varepsilon^{4} & \varepsilon^{6} & \varepsilon^{8}%
\end{array}
\right)  ,\text{ }\lambda_{DER}\sim\left(
\begin{array}
[c]{ccc}%
1 & \varepsilon^{2} & \varepsilon^{3}\\
\varepsilon^{2} & \varepsilon^{3} & \varepsilon^{4}\\
\varepsilon^{3} & \varepsilon^{4} & \varepsilon^{5}%
\end{array}
\right)  .
\end{equation}
As estimated in \cite{Heckman:2008qa}, crude scaling arguments relate $\varepsilon$ to the GUT fine structure constant as $\varepsilon \sim \sqrt{\alpha_{GUT}}$. One of our aims in this section will be to see how similar structures appear in the non-commutative theory we have studied.

To summarize, counting the number of powers of $z$ and $\bar{z}$ in a physical basis for the wave-functions constrains the form of possible higher order
corrections to the structure of the Yukawa couplings. Note, however, that
counting powers of $z$ and $\bar{z}$ certainly obscures the overall
holomorphy of the superpotential. Indeed, one of the aims of this paper has
been to develop a manifestly holomorphic formulation of Yukawa couplings.

\subsection{Comparison with the Non-Commutative Gauge Theory}

In the previous section we reviewed the argument that a generic background
flux will alter the internal profile of matter field
wave-functions, and thus the Yukawa interactions as well. A more precise
analysis reveals, however, that background gauge field fluxes alone
do not disort the structure of the Yukawas. On the other hand, we have seen in this paper that background
$H$-fluxes do distort the structure of the Yukawas. Indeed, at the level
of matching tensor indices, both $H$-fluxes and gradients of the gauge field
strength with one holomorphic and two anti-holomorphic indices given by:%
\begin{equation}
H_{i{\bar j \bar k}}\text{, }\nabla_{i}F_{\bar j \bar k}\text{, }\nabla
_{\bar{i}}F_{j\bar{k}}%
\end{equation}
are of the general form required for the FLX\ and DER\ expansions.

The contributions from the $H$-flux and the gauge field strength are actually
closely related. For example, in the simplified case of perturbative type
IIB\ vacua, the DBI\ action for a single D7-brane contains the gauge field
strength $F$ in tandem with the NS$\ B$-field through the combination
$F+B$. Indeed, in this simplified case, one of the equations of motion for the
D7-brane theory is:%
\begin{equation}
F_{{\bar j\bar k}}+B_{{\bar j\bar k}}=0
\end{equation}
for the $(0,2)$ component of the gauge field strength and the background
$B$-field. Taking the gradient of this equation, we then obtain a relation
between the gauge field strength and the $H$-flux in directions along the
D7-brane:%
\begin{equation}
\nabla_{i}F_{{\bar j\bar k}}+H_{i{\bar j\bar k}}=0.
\end{equation}
From this perspective, it is now immediate that the presence of the background
$H$-flux can simply be viewed as a deformation of the ordinary Hermitian
Yang-Mills equations by higher dimension operators. Although a full discussion
is beyond the scope of this paper, it would be interesting to study in more
general terms higher dimension operator contributions to the superpotential.
In particular, such an analysis likely admits an interpretation in terms of
Massey products.

\subsection{Scaling Estimates}

The structure of Yukawa matrices found in section 5 are characterized in terms
of $\hbar_{\theta}$, the quantum expansion parameter
of the non-commutative deformation and $\Phi_{0}$, the constant part of the trace of the $(2,0)$
form in a sub-block of the matrix $\phi^{(0)}$. We now estimate the
scaling behavior of these parameters. In the next subsection we shall use
these estimates to compare the hierarchical expansions found in previous
sections with those of the FLX\ and DER\ expansions.

To estimate the scaling behavior of $\hbar_{\theta}$, we first recall the
primary mass scales which enter into F-theory GUTs. First, there is the
characteristic scale of the complex surface $S$, which is given by a mass
scale $M_{GUT}$ such that $M_{GUT}^{4}\sim1/$Vol$(S)$. In addition, there is a
characteristic mass scale associated with stringy modes, which we shall refer
to as $M_{\ast}$. For example, the units of gauge field flux naturally scale
as $F\sim M_{GUT}^{2}$, while the three-form $H$-flux which is not localized
on a particular seven-brane more naturally scales as $H\sim M_{\ast}%
^{3}$. The fine structure constant of the GUT is given in terms of
$M_{GUT}$ and $M_{\ast}$ as \cite{BHVII}:%
\begin{equation}
\frac{M_{GUT}^{4}}{M_{\ast}^{4}}\sim\alpha_{GUT}\sim\frac{1}{25}.
\end{equation}
Using these two basic mass scales, we now proceed to estimate the scaling of
$\hbar_{\theta}$ and $\Phi_{0}$.

First consider the quantum expansion parameter $\hbar_{\theta}$.
In units where $\hbar_{\theta}$ is set to $1$, an estimate for the size of the
non-commutative deformation can be deduced by computing the overall scaling of the Poisson
bi-vector $\theta$. Since a pointlike probe D3-brane can detect the presence
of this parameter, it follows that the overall scaling of $\theta$ is fully
determined by the index structure of $\theta$. In particular, it follows that
although $\theta$ will typically be a non-trivial function of the internal
coordinates of the complex surface, this further functional dependence can
effectively be viewed as a dimensionless contribution. Thus, it is enough to
evaluate the overall size of the parameter $\theta$ based on its index structure.

As shown in section 4, the parameter $\theta$ is defined by its relation to
the background $H$-flux through the equation:%
\begin{equation}
\nabla_{i}\theta^{jk}=H_{i}^{\text{ \ }jk}=g^{j\overline{j}}g^{k\overline{k}%
}H_{i\overline{jk}}.
\end{equation}
Now, although the $H$-flux is naturally quantized in units suited to $M_{\ast
}$, the metric on the complex surface will instead scale as an appropriate
power of $M_{GUT}$. Moreover, because the volume of the complex surface is
given as:%
\begin{equation}
\text{Vol}(S)=\underset{S}{\int}\sqrt{g}\text{,}%
\end{equation}
it follows that each inverse power of the metric contributes a factor of
$M_{GUT}^{2}$ to the overall scaling of $\theta$. Finally, because the
gradient $\nabla_{i}$ is again measured in closed string units, we conclude
that the only $M_{GUT}$ scaling is, in dimensionless units:%
\begin{equation}
\theta^{jk}\sim\frac{M_{GUT}^{4}}{M_{\ast}^{4}}\sim\alpha_{GUT}\sim
\varepsilon^{2},
\end{equation}
where in the final estimate we have introduced an expansion parameter
$\varepsilon$ defined as in \cite{Heckman:2008qa}.

Next consider the scaling of the parameter $\Phi_{0}$, the constant part of
the trace of a sub-block of the $(2,0)$ form, or more abstractly, by its trace
in the universal enveloping algebra. It might at first appear
that $\Phi_{0}$ should scale as one power of $M_{GUT}$,
associated with the scaling of a single field. Writing $\varphi = \phi dx \wedge dy$ on a patch, since $||\phi||^2$ scales inversely with the volume, $\Phi_{0}$ scales as $M_{GUT}^2$. This is also in accord with the fact that $\varphi$ has two indices along the
GUT seven-brane, scaling in the same way as the gauge field
flux in the internal directions of the seven-brane theory, namely, as $M_{GUT}^2$. As a unitless expansion parameter, $\Phi_{0}$ therefore scales as:%
\begin{equation}
\Phi_{0}\sim\frac{M_{GUT}^{2}}{M_{\ast}^{2}}\sim\sqrt{\alpha_{GUT}}%
\sim\varepsilon.
\end{equation}
Summarizing the analysis just presented, the expansion parameters of the
non-commutative gauge theory are:%
\begin{equation}
\hbar_{\theta}\sim\varepsilon^{2}, \;\; \Phi_{0}\sim\varepsilon.%
\end{equation}

\subsection{Matching to the Non-Commutative Gauge Theory}

We now compare the hierarchical Yukawa matrices found in section 5 with those
of the FLX\ and DER expansions. We find that the most phenomenologically
attractive Yukawa matrices originate from configurations where the holomorphic
bi-vector $\theta$ vanishes to first order at the interaction point. In this specific case,
the types of Yukawa structures for a $g$ generation model are of the form:%
\begin{equation}
\lambda_{NC}\sim\left(
\begin{array}
[c]{ccccc}%
1 & \hbar_{\theta} & \hbar_{\theta}^{2} & \hbar_{\theta}^{3} & ...\\
\hbar_{\theta} & \hbar_{\theta}^{2} & \hbar_{\theta}^{3} & \hbar_{\theta}^{4}
& ...\\
\hbar_{\theta}^{2} & \hbar_{\theta}^{3} & \hbar_{\theta}^{4} & \hbar_{\theta
}^{5} & ...\\
\hbar_{\theta}^{3} & \hbar_{\theta}^{4} & \hbar_{\theta}^{5} & \hbar_{\theta
}^{6} & ...\\
... & ... & ... & ... & ...
\end{array}
\right)  +\Phi_{0}\left(
\begin{array}
[c]{ccccc}%
0 & 0 & \hbar_{\theta} & \hbar_{\theta}^{2} & ...\\
0 & \hbar_{\theta} & \hbar_{\theta}^{2} & \hbar_{\theta}^{3} & ...\\
\hbar_{\theta} & \hbar_{\theta}^{2} & \hbar_{\theta}^{3} & \hbar_{\theta}^{4}
& ...\\
\hbar_{\theta}^{2} & \hbar_{\theta}^{3} & \hbar_{\theta}^{4} & \hbar_{\theta
}^{5} & ...\\
... & ... & ... & ... & ...
\end{array}
\right)  +\Phi_{0}^{2}\left(
\begin{array}
[c]{ccccc}%
0 & 0 & 0 & 0 & ...\\
0 & 0 & 0 & \hbar_{\theta}^{2} & ...\\
0 & 0 & \hbar_{\theta}^{2} & \hbar_{\theta}^{3} & ...\\
0 & \hbar_{\theta}^{2} & \hbar_{\theta}^{3} & \hbar_{\theta}^{4} & ...\\
... & ... & ... & ... & ...
\end{array}
\right)  +....\label{NCYUKYUK}%
\end{equation}
By inspection of the above matrices, we conclude that the exact form of the
FLX expansion is reproduced by the $\Phi_{0}$ independent contribution to
equation (\ref{NCYUKYUK}). Indeed, setting $\hbar_{\theta}\sim\varepsilon^{2}%
$ yields:%
\begin{equation}
\lambda_{\hbar}\equiv\lambda_{NC}|_{\Phi_{0}=0}\sim\lambda_{FLX}\sim\left(
\begin{array}
[c]{ccccc}%
1 & \varepsilon^{2} & \varepsilon^{4} & \varepsilon^{6} & ...\\
\varepsilon^{2} & \varepsilon^{4} & \varepsilon^{6} & \varepsilon^{8} & ...\\
\varepsilon^{4} & \varepsilon^{6} & \varepsilon^{8} & \varepsilon^{10} & ...\\
\varepsilon^{6} & \varepsilon^{8} & \varepsilon^{10} & \varepsilon^{12} &
...\\
... & ... & ... & ... & ...
\end{array}
\right)  .
\end{equation}
The form of the FLX expansion requires that terms independent of $\Phi_{0}$
dominate over those which have some $\Phi_{0}$ dependence. Indeed, as argued
in \cite{Heckman:2008qa}, the FLX\ expansion is expected to dominate in
Yukawas involving matter fields with large couplings to background fluxes.

The DER expansion is expected to dominate in those Yukawas involving matter
fields with smaller hypercharges. In the present context, it is natural to
expect that this milder type of coupling is captured through the power series
in $\Phi_{0}$. Working to leading order in this expansion, the form of
$\lambda_{NC}$ given by equation (\ref{NCYUKYUK}) is:%
\begin{equation}
\lambda_{\Phi}\sim\left(
\begin{array}
[c]{ccccc}%
1 & \hbar_{\theta} & \Phi_{0}\cdot\hbar_{\theta} & \Phi_{0}\cdot\hbar_{\theta
}^{2} & ...\\
\hbar_{\theta} & \Phi_{0}\cdot\hbar_{\theta} & \Phi_{0}\cdot\hbar_{\theta}^{2}
& \Phi_{0}^{2}\cdot\hbar_{\theta}^{2} & ...\\
\Phi_{0}\cdot\hbar_{\theta} & \Phi_{0}\cdot\hbar_{\theta}^{2} & \Phi_{0}%
^{2}\cdot\hbar_{\theta}^{2} & \Phi_{0}^{2}\cdot\hbar_{\theta}^{3} & ...\\
\Phi_{0}\cdot\hbar_{\theta}^{2} & \Phi_{0}^{2}\cdot\hbar_{\theta}^{2} &
\Phi_{0}^{2}\cdot\hbar_{\theta}^{3} & \Phi_{0}^{3}\cdot\hbar_{\theta}^{3} &
...\\
... & ... & ... & ... & ...
\end{array}
\right)  \sim\left(
\begin{array}
[c]{ccccc}%
1 & \varepsilon^{2} & \varepsilon^{3} & \varepsilon^{5} & ...\\
\varepsilon^{2} & \varepsilon^{3} & \varepsilon^{5} & \varepsilon^{6} & ...\\
\varepsilon^{3} & \varepsilon^{5} & \varepsilon^{6} & \varepsilon^{8} & ...\\
\varepsilon^{5} & \varepsilon^{6} & \varepsilon^{8} & \varepsilon^{9} & ...\\
... & ... & ... & ... & ...
\end{array}
\right)  .
\end{equation}
Although similar, the precise structure of the expansion $\lambda_{\Phi}$
deviates from the DER expansion by a small amount. Indeed, restricting to the
case of a three generation model, we see that by comparison, the three entries
in the lower righthand corner of the $3\times3$ matrix differs by a factor of
$\varepsilon$:%
\begin{equation}
\lambda_{\Phi}\sim\left(
\begin{array}
[c]{ccc}%
1 & \varepsilon^{2} & \varepsilon^{3}\\
\varepsilon^{2} & \varepsilon^{3} & \varepsilon^{5}\\
\varepsilon^{3} & \varepsilon^{5} & \varepsilon^{6}%
\end{array}
\right)  \text{, }\lambda_{DER}\sim\left(
\begin{array}
[c]{ccc}%
1 & \varepsilon^{2} & \varepsilon^{3}\\
\varepsilon^{2} & \varepsilon^{3} & \varepsilon^{4}\\
\varepsilon^{3} & \varepsilon^{4} & \varepsilon^{5}%
\end{array}
\right)  .
\end{equation}
Let us stress that there could in principle be additional contributions to the
Yukawa matrices which might reproduce exactly the form of the DER expansion. As
an example, one might consider more elaborate contributions from higher order terms in the series expansions:
$$
\theta(x,y) = \sum_{m_1+m_2 > 0} \theta_{m_1,m_2} x^{m_1} y^{m_2},
$$
and:
$$
\Phi(x,y) = \Phi_0 + \sum_{m_1+m_2 > 0} \Phi_{m_1,m_2} x^{m_1} y^{m_2}.
$$
Our aim here has been to explore the most generic situation consistent with
a $\theta$ which vanishes to first order. It would be interesting to see how much
``fine-tuning'' of higher order corrections is necessary to reproduce precisely
the form of the DER expansion just from the non-commutative deformation already studied.

Indeed, given the quite general arguments
provided in \cite{Heckman:2008qa}, it is reasonable to expect that additional
corrections to the structure of the Yukawa will likely reproduce precisely
the structure of the DER expansion. For example, there will typically be
higher order corrections to the relation $\nabla \theta = H$ found for weak $H$-fluxes.
In the effective field theory, such corrections constitute the presence of additional
higher dimension operators, which might lead to a milder hierarchy, in line with the estimates of
\cite{Heckman:2008qa}.

There is also a broader class of non-commutative deformations which one might consider.
Indeed, besides the non-commutative deformation
in the local coordinates $x$ and $y$, one might also consider a non-commutative deformation of the form:
$$
[x,y] = \hbar_{S}\theta^{xy}, \;\; [y,z] = \hbar_{\bot}\theta^{yz}, \;\; [z,x] = \hbar_{\bot}\theta^{zx}
$$
where $z$ denotes a local coordinate parameterizing directions normal to the seven-brane, and we
have distinguished the expansion parameters $\hbar_{S}$ and $\hbar_{\bot}$ for Poisson bi-vectors with respectively
both legs, or one leg along the seven-brane wrapping the complex surface $S$. Although
it is beyond the scope of the present paper, we now briefly sketch in quite heuristic terms how a
computation of Yukawas in this setting might go. The effects of this deformation are most readily analyzed
as a non-commutative deformation of holomorphic Chern-Simons
theory. In this language, the configuration of seven-branes and matter curves
are specified by expanding about a background field configuration which satisfies the BPS equations of motion. Note that
just as the location of the matter curves in the seven-brane theory is no longer
sharply defined once $\hbar \neq 0$, so too will the location of the seven-branes become fuzzy. This is in accord with the co-isotropic condition for
seven-branes studied in \cite{kapustin}. Even so, just as for the seven-brane theory we have analyzed, the theory still
contains localized zero modes, and so an appropriate cohomology theory can be developed
in this case as well. The holomorphic dependence of the zero modes will now contain
dependence on the local coordinates of the seven-brane, $x$ and $y$, as well as $z$,
the coordinate normal to the seven-brane. It is therefore quite tempting to simply
take the formal structure of the Yukawa matrix $\lambda_{\hbar}$, and replace
$\hbar_{\theta}^{n}$ by either $\hbar_{S}^{n}$ or $\hbar_{S} \cdot \hbar_{\bot}^{n-1}$. Assuming
that the number of indices along the GUT seven-brane dictates the overall scaling of the Poisson bi-vector, setting
$\hbar_{S} \sim \varepsilon^2$ and $\hbar_{\bot} \sim \varepsilon$ would now
appear to provide an exact match respectively to the FLX and DER
expansions. It would be interesting to perform a more precise computation and
estimate of the resulting Yukawas, along similar lines to what has been developed in this paper.

\subsection{Masses and Mixing Angles with $\lambda_{\hbar}$ and $\lambda_{\Phi}$}

Restricting now to the Yukawa structures $\lambda_{\hbar}$ and $\lambda_{\Phi}$ found
in this paper, we now ask whether these structures reproduce the main features of
the estimates for the mixing angles and masses given in \cite{Heckman:2008qa}.

To compare with these estimates, we switch
to a basis of states ordered from lightest to heaviest in mass, so that for
example, the $(3,3)$ entry of $\lambda_{DER}$ is of order one, while the
$(1,1)$ entry is of order $\varepsilon^{5}$. In \cite{Heckman:2008qa}, the
up-type quark Yukawa was identified with the FLX expansion, while the
down-type Yukawa was matched to the DER expansion. Identifying the up-type
quark Yukawa matrix with the $\lambda_{\hbar}$ matrix and that of the
down-type Yukawa matrix by $\lambda_{\Phi}$, the three generation
CKM\ matrix is of the form:%
\begin{equation}
V_{CKM}\sim\left(
\begin{array}
[c]{ccc}%
1 & \varepsilon & \varepsilon^{3}\\
\varepsilon & 1 & \varepsilon^{2}\\
\varepsilon^{3} & \varepsilon^{2} & 1
\end{array}
\right)  .
\end{equation}
As in \cite{Heckman:2008qa}, it is interesting to estimate the form of the
CKM\ matrix in the case of a $g$-generation model for $g\geq3$ in which the
three generations of the Standard Model are the lightest. We find that the
CKM\ matrix of the Standard Model alternates between the structure
for odd and even generations:%
\begin{equation}
V_{CKM}^{g\text{ odd}}\sim\left(
\begin{array}
[c]{ccc}%
1 & \varepsilon & \varepsilon^{3}\\
\varepsilon & 1 & \varepsilon^{2}\\
\varepsilon^{3} & \varepsilon^{2} & 1
\end{array}
\right)  \text{, }V_{CKM}^{g\text{ even}}\sim\left(
\begin{array}
[c]{ccc}%
1 & \varepsilon^{2} & \varepsilon^{3}\\
\varepsilon^{2} & 1 & \varepsilon\\
\varepsilon^{3} & \varepsilon & 1
\end{array}
\right)  .
\end{equation}
In other words, the expected hierarchy in a three generation model is in
better agreement with observation than what is expected in the four generation model.

Next consider the masses of associated with these Yukawa structures. Since the
up-type quarks and charged leptons both descend from the FLX expansion, which
is reproduced by $\lambda_{\hbar}$, the estimates for these masses are the
same as in \cite{Heckman:2008qa}. On the other hand, the masses from
$\lambda_{\Phi}$ are slightly different from those of the DER\ expansion:%
\begin{align}
NC &  :m_{d}:m_{s}:m_{b}\sim\varepsilon_{d}^{6}:\varepsilon_{d}^{3}:1\\
DER &  :m_{d}:m_{s}:m_{b}\sim\varepsilon_{d}^{5}:\varepsilon_{d}^{3}:1.
\end{align}
Fitting to the masses of the down type quarks, this leads to a slightly higher
estimate for the value of $\varepsilon_{d}$ when compared to what is expected
from the DER expansion. Nevertheless, the resulting value of $\varepsilon_{d}$
is still on the order of $\sqrt{\alpha_{GUT}}\sim0.2$.

\section{Conclusions}

Yukawa couplings provide an important link between the physics of the Standard
Model and that of possible ultraviolet completions. In this paper, we have
seen that computations of Yukawa couplings in the context of F-theory with
background gauge field and $H$-fluxes activated admit a rather interesting
local geometric formulation near the intersection point of seven-branes.
As we have shown in this paper, Yukawa couplings map to a residue integral computation defined in a neighborhood containing the Yukawa enhancement point. Although gauge field fluxes alone do not alter the structure of the Yukawas, background $H$-fluxes do deform the structure of the Yukawa couplings. Moreover, when $H$-fluxes are turned on, there is a natural formulation of the effective superpotential in terms of a non-commutative space with non-commutativity fixed by the strength of the $H$-flux. We have also found evidence that background fluxes can generate flavor hierarchies in F-theory GUTs of the type proposed in
\cite{Heckman:2008qa} provided we identify the local profile of the three-form
field with the three-form flux $H$.

There are a number of issues which need further study. On the theoretical
side, it would be important to better understand the non-commutative theory
for $E$-type groups. Even though in principle this can be done along the lines
suggested in \cite{wess} by allowing all fields to take values in the
universal envelopping algebra, from the viewpoint of F-theory it is clear that
there are some subtleties to be resolved. For example, in general F-theory
compactifications, $H$ and consequently $\theta$ as well will undergo $SL(2,%
\mathbb{Z}
)$ monodromies near such enhancement points. It would be important to better
understand this feature, especially in view of the fact that F-theory
GUTs demand $E$-type enhancement points.

In more realistic settings, the configuration of seven-branes will contain
monodromies. It is clearly important to study the effects of such monodromies
on the form of the superpotential, and the interplay between such geometric
ingredients and fluxes. For example, in order to obtain the requisite flavor
hierarchy from $H$-flux, we assumed that $\theta$ should vanish to first order at the Yukawa enhancement point. It would be interesting to understand how this may be related to seven-brane monodromies.

The study of non-commutative deformations of topological strings related to
the extended moduli space is in its infancy. We have already uncovered a very
rich structure in the disk amplitudes for a deformation in this generalized
moduli space. It would be very interesting to study this issue more
systematically, as well as in a more general setup. Along these lines, it
would be interesting to study the structure of Yukawa couplings expected directly
in terms of the non-commutative deformation of holomorphic Chern-Simons theory.

Finally, from a phenomenological point of view, the contribution from the non-commutative deformation of the F-terms can be viewed as a specific class of higher dimension operators. It would be interesting to determine the most general class of such higher dimension operators, and the resulting flavor hierarchies expected.

\section*{Acknowledgements}

We thank A. Tavanfar for discussions and collaboration at an early stage of
this project. We also thank F. Denef, O. Lunin, A. Neitzke, V. Pestun, M. Rocek, A. Tomasiello and M. Wijnholt for helpful discussions. MC, JJH and CV would also like to thank the Seventh Simons Workshop in Mathematics and Physics for
hospitality while some of this work was performed. The work of MC is supported by the Netherlands Science Organisation (NWO). The work of JJH and CV was
supported in part by NSF grant PHY-0244821. The work of JJH is also supported
by NSF grant PHY-0503584.

\appendix

\section{Calculation of the Commutative Residue}
\label{Calculation of the Residue Formula}

In this Appendix we compute the Yukawa coupling in the commutative theory between localized zero modes of the form (\ref{z_modes_Dexpression}) and show that the answer is given by the residue formula (\ref{residue_formula_commutative}). Starting from
\bea
W_{\text{Yuk}} &=& \int_U \Tr(\Psi^{(1)}\wedge \Psi^{(1)}\wedge \Psi^{(1)}) \\ \notag
&=& \int_U \Psi_{\a_1}\wedge \Psi_{\a_2}\wedge \bar D  \Big(\frac{\fione_{\a_3}-h_{\a_3}}{\a_3(\finull)}\Big)
+\int_U  h_{\a_3}\, \Psi_{\a_1}\wedge \Psi_{\a_2}\wedge \, dx\wedge dy\;,
\eea
we will first show that we can discard the first term.
Using the Leibniz rule (\ref{Leibnitz}) and the fact that $\bar D =\bar\pa$ on a gauge singlet to write it as a boundary term
$$
\int_{\pa U}  \Psi_{\a_1}\wedge \Psi_{\a_2}\wedge   \Big(\frac{\fione_{\a_3}-h_{\a_3}}{\a_3(\finull)}\Big)\;.
$$
From the fact that $\Psi_{\a_1}$ and $\Psi_{\a_2}$ are localized on two distinct, transversely intersecting matter curves which only intersect at the point our patch $U$ is chosen to enclose and in particular do not intersect at the boundary of the patch, we conclude that this surface term does not contribute to the Yukawa integral.

This statement can be made more precise in the following way.
As we argued in section \ref{Zero Modes Localized on Curves}, the superpotential and in particular the Yukawa coupling should not depend on which gauge orbit, or equivalently which representative in the cohomology,  one chooses for the wave-function. So let us now choose the zero modes
\be
\Psi^{(1)}_\a X_\a= \bar D \Big(\frac{\fione_\a-h_\a}{\a(\finull)}\, X_\a\Big) + \hat\k \wedge h_\a X_\a\quad,\quad \fione_\a\lvert_{\a(\finull) = 0} \, =h_\a,
\ee
with the function $\fione_\a$ localized in a tube of radius $\r_\a$ surrounding the matter curve $\Sigma_\a$:
\be\label{tubular_support}
\fione_\a = 0 \quad\text{when}\quad |\a(\finull)| > \r_\a \;.
\ee
Let $K$ be any convex compact set in $\C^2$ with smooth boundary containing the triple intersection point
and let $K(R)$ be $K$ rescaled by $R$. Take our patch to be $U=K(R \gg \rho_\a )$.
Then from the fact that $T_{\a_1} \cap T_{\a_2}\cap \pa U = \emptyset$, where $T_{\a}$ is the tube centering around the matter curve $\Sigma_{\a}$ with radius $\r_\a$,
 we see that this boundary term indeed vanishes.

Let us now compute the remaining term
\bea\notag
W_{\text{Yuk}} &=&
\int_U  h_{\a_3}\, \Psi_{\a_1}\wedge \Psi_{\a_2}\wedge \, dx\wedge dy \\&=& \int_U \; h_{\a_3}\, \bar \pa_{A^{(0)}}  \Big(\frac{\fione_{\a_1}-h_{\a_1}}{\a_1(\finull)}\Big)\wedge \bar  \pa_{A^{(0)}}  \Big(\frac{\fione_{\a_2}-h_{\a_2}}{\a_2(\finull)}\Big) \, dx\wedge dy\;,
\eea

Since $h_\a$ is holomorphic, we can also write this term as a boundary term
\bea\notag
&&\int_{\pa U} \;h_{\a_3}   \Big(\frac{\fione_{\a_1}-h_{\a_1}}{\a_1(\finull)}\Big)\wedge \bar  \pa_{A^{(0)}}  \Big(\frac{\fione_{\a_2}-h_{\a_2}}{\a_2(\finull)}\Big)  \wedge dx\wedge dy\\&=&
\int_{\pa U \cap T_{\a_2}} \;h_{\a_3}   \Big(\frac{\fione_{\a_1}-h_{\a_1}}{\a_1(\finull)}\Big)\wedge \bar  \pa_{A^{(0)}}  \Big(\frac{\fione_{\a_2}-h_{\a_2}}{\a_2(\finull)}\Big) dx\wedge dy\\ \notag
&=&
\int_{\pa U \cap T_{\a_2}} \;-   \frac{h_{\a_1}\,h_{\a_3}}{\a_1(\finull)}\wedge \bar  \pa_{A^{(0)}}  \Big(\frac{\fione_{\a_2}-h_{\a_2}}{\a_2(\finull)}\Big) dx\wedge dy
\;,
\eea
where we have again used the support property of $\fione_{\a_1}$ and $\fione_{\a_2}$.

From the holomorphicity of $h_\a$ and $\finull$ we can now perform another integration by parts and write
\bea
W_{\text{Yuk}}  = \int_{\pa U \cap \pa T_{\a_2}} \frac{h_{\a_1}\,h_{\a_2}\,h_{\a_3}}{\a_1(\finull) \a_2(\finull)} dx\wedge dy
\eea

This is a residue integral, and in particular $\,\pa U \cap \pa T_{\a_2}\,$ is diffeomorphic to a product of two circles surrounding the intersection point. Hence the final answer can be written as equation (\ref{residue_formula_commutative}).

\section{The $\star$ and $\ost$ Product}

The precise definition of $\ost$ is given as a natural extension of the $\star$ product for functions to the case of differential forms. Both operations are defined in terms of an expansion in successive powers of $\hbar_{\theta}$. When the Poisson bi-vector is constant, the $\star$-product between two functions is simply the Moyal product
\be\label{Moyal}
f\star g = e^{\hbar_{\theta}\,\theta\, \pa_x \wedge \pa_y  } \,\big(f\otimes g\big) := \sum_{n=0}^\inf \sum_{k=0}^n \frac{(\hbar_{\theta} \theta)^n }{k!(n-k)!} \, (-1)^k\, \big(\pa_x^{n-k}\pa_y^k f \big)\, \big( \pa_x^{k}\pa_y^{n-k} g\big)\;.
\ee

The form of the $\star$ product is more involved when the Poisson bi-vector is not constant. To illustrate this point, now consider the form of $\star$ in a neighborhood $U$ of the classical Yukawa point, which we take to lie at $x = y = 0$, in local coordinates $x$ and $y$ of $U$. When $\theta$ does not vanish at the origin, the Darboux theorem ensures we can find local coordinates $X,Y$ such that $\hat \theta = \pa_X\wedge \pa_Y$. The situation is more complicated when $\theta$ does vanish at the origin. Nevertheless we still have the following result: in the neighborhood $U$ of the origin, denoting $D\equiv (\theta)$ as the divisor of vanishing $\theta$, then on $U\setminus D$ there exist two holomorphic vector fields $V_1$, $V_2$ such that
\be\label{def_V}
\hat \theta = \theta(x,y) \pa_x\wedge \pa_y = V_1 \wedge V_2 \quad,\quad [V_1,V_2] =0\;,
\ee
given that either $\small{(1)} \;\theta(0,0)\neq 0$, or (2) the divisor $D$ has support on a smooth curve passing through the origin, or (3) the divisor $D$ has support on two smooth component curves meeting transversely at the origin. The simplest way to convince oneself of the above statement is to observe that the above three classes of situations correspond respectively to the following three classes of Poisson bi-vectors (1) $V_1=\pa_x, V_2 = \pa_y$, (2)
$V_1=x \pa_x, V_2 = \pa_y$, (3)
$V_1=x\pa_x, V_2 = y\pa_y$.

The $\star$-product, in this case, is the simple generalization of the Moyal product (\ref{Moyal}):
\be\label{starproduct}
f\star g = e^{\hbar_{\theta} V_1 \wedge V_2} =  \sum_{n=0}^\inf \sum_{k=0}^n \frac{ (-1)^k \hbar_{\theta}^n }{k!(n-k)!} \, \big(V_1^{n-k}V_2^k f \big)\, \big( V_1^{k}V_2^{n-k} g\big)\;.
\ee
Recall that a theorem by Kontsevich \cite{kontsevich1} states that the requirement for the $\star$-product to be associative and having the leading behavior
$$
f\star g = fg + \hbar_{\theta}\, \theta\, (\pa_x f \pa_y g - \pa_y f \pa_x g) + {\cal O}(\hbar_{\theta}^2)
$$
determines the $\star$-product as a formal series in $\hbar_{\theta}$ uniquely for a given Poisson bi-vector $\theta$. And hence the $\star$-product (\ref{starproduct}) we defined above is in fact unique.

We next define the $\ost$-product, the non-commutative deformation of the $\wedge$-product, acting on forms.
Since we are concerned with non-commutativity in holomorphic coordinates only, the $\ost$-products between a $(n_1,m_1)$- and a $(n_2,m_2)$-form take the same form for all $m_1,m_2 = 0,1,...$.
The natural generalization of the non-commutative product from the space of functions to that of all forms $\oplus_{(p,q)} \L^{(p,q)}(U)$ on the neighborhood $U$ is to ``covariantize'' the product by writing forms in the basis $\varpi_1, \;\varpi_2$ which are meromorphic closed one-forms dual to the vectors $V_1,\;V_2$ defined above, and then taking the $\star$-product between the coefficients. This basis for one-forms and vectors can be thought of as the analogue of the ``vielbein'' basis but now defined by the Poisson structure.

To discuss the deformation of the field theory with superpotential given by (\ref{superpotential}), we need to define the $\ost$-products among $(0,*)$- and $(2,*)$-forms. For $(0,*)$-forms $f$ and $g$, we have:
\begin{equation}
f\ost g = f \star g.
\end{equation}
Next consider the $\ost$ product between a $(2,*)$ form,
$\Psi = \psi \, dx \wedge dy$ and a $(0,*)$ form $f$. To follow the covariantization
procedure described above, note that when $\Psi$ is for example a $(2,0)$ form,
$\psi$ is a section of the canonical bundle $K_S = (\Lambda^{2}TS)^{\ast}$. More generally,
note that although $\psi$ and $\theta$ both transform non-trivially under a change of
coordinates, their product $\theta \psi$ can be treated as a $(0,*)$-form.
The $\ost$ product is therefore given as:
\bea\notag
f \ost (\psi\,dx\wedge dy)
 &=&  f\star (\theta \,\psi) \,\frac{dx\wedge dy}{\theta}\\ \notag\label{ostar}
   (\psi\,dx\wedge dy) \ost f
& =&   (\theta \,\psi) \star f\; \frac{dx\wedge dy}{\theta}\;.
 \eea
We shall also sometimes write:
\bea\notag
f \ost \psi
 &=&  (f\star (\theta \,\psi)) \,\frac{1}{\theta}\\ \notag
   \psi \ost f
& =&   ((\theta \,\psi) \star f)\,\frac{1}{\theta}\;.
 \eea

The $\ost$-product between differential forms satisfies the following important properties:
\begin{enumerate}
\item{Associativity}
\item{The $\bar\pa$-chain rule}
$$
\bar\pa (F \ost F') = \bar\pa F \ost F' +(-1)^{\text{deg}F} F  \ost\bar\pa F'\quad \quad\text{for $(p,q)$-forms }F,F'
$$
\item{Commutativity Upon Integration}
\be\label{SWconditionAPP}
\int_S \Tr \big( F\ost F' \big) =\int_S \Tr \big( F'\ost F \big) \;,
\ee
\end{enumerate}
for forms $F$ and $F'$ which satisfy an appropriate notion of localization. These properties are in turn crucial in order for the non-commutative deformation of the seven-brane superpotential to be well-defined.

\section{Proof of Commutativity Upon Integration}
\label{Proof of Commutativity Upon Integration}

In this Appendix we prove the claim mentioned in section 5, line (\ref{SWcondition}) and in Appendix B, line (\ref{SWconditionAPP}), namely that the $\ost$-product we defined in (\ref{ostar}) satisfies the property:
\be
\int_S \Tr \big( F\ost F' \big) =\int_S \Tr \big( F'\ost F \big) \;,
\ee
for forms $F$ and $F'$ which satisfy an appropriate notion of localization.

Note that for our present purposes, it is enough to show that for a $(0,2)$-form $f$ and a $(2,0)$-form $\Psi = \psi \, dx \wedge dy$ which satisfy the localization condition:
$$
\pa_x^{m_1}\pa_y^{m_2} f \,\pa_x^{n_1}\pa_y^{n_2} \psi \to 0 \quad \text{for all }m_1,m_2,n_1,n_2 >0
$$
at the boundary of the patch $U$, that the following property holds:
\be
\int_U f\ost\big(\psi \, dx\wedge dy\big)=\int_U \big(\psi \, dx\wedge dy\big) \ost f \;.
\ee

To prove this, first define
$$
\hat V_i = V_i + V_i \log \th\quad,\quad i=1\;,
$$
and then write
\bea\notag
&&f\ost \big(\psi \,dx\wedge dy \big)- \big(\psi \,dx\wedge dy \big)\ost f\\ \notag&=&
-2 \sum_{n=0}^\inf \hbar_\th^{1+2n} \sum_{k=0}^{1+2n} \frac{(-1)^k }{k! (1+2n-k)!}\, \big(V_1^k V_2^{1+2n-k} f \big)
\big(\hat V_1^{1+2n-k} \hat V_2^{k} \psi \big) dx\wedge dy\\ \notag&=&
-2\sum_{n=0}^\inf \frac{\hbar_\th^{1+2n}}{(2n+1)!} \sum_{k=0}^{2n} (-1)^k \frac{(2n)!}{k! (2n-k)!} \big\{V_2 (V_1^{2n-k}V_2^kf)\hat V_1 (\hat V_1^{k}\hat V_2^{2n-k}\psi)\\\notag
&&  - V_1 (V_1^{2n-k}V_2^kf)\hat V_2 (\hat V_1^{k}\hat V_2^{2n-k}\psi)\big\} \, dx\wedge dy\;.
\eea

Note that each summand in the above expression can be written as
$$
(V_2 \til f \hat V_1 \til \psi - V_1 \til f \hat V_2 \til \psi) \,dx\wedge dy
$$
for some $\til f$ and $\til \psi$. It is therefore enough to show that each
of these terms corresponds to an exact differential form. To this end, note that:
\bea \notag
&& (V_2 \til f \hat V_1 \til \psi - V_1 \til f \hat V_2 \til \psi) \,dx\wedge dy\\ \notag
&=& - \big\{( V_1 \til f V_2 \til \psi - V_2 \til f V_1 \til \psi)  + \frac{\til \psi}{\th}(V_1 \til f V_2 \th - V_2 \til f V_1 \th) \big\} \, dx\wedge dy\\ \notag
&=&- \big(\pa_x \til f \pa_y (\th \til \psi) -\pa_y \til f \pa_x (\th \til g) \big)\, dx\wedge dy\\ \notag
&=& -d\til f\wedge d(\til \psi \th)\;.
\eea
This proves our claim.

\end{document}